\documentclass [12pt]{article}
\usepackage {graphicx}
\usepackage {amssymb}
\usepackage {amsmath}
\usepackage {longtable}
\title{Evolution of vacancy pores in bounded particles.}
\author{$^{1,2}$\textbf{V.V. Yanovsky}, $^1$\textbf{M.I. Kopp}, $^1$\textbf{M. A. Ratner}}

\begin{document}
\maketitle

$^{1}$ \textit{Institute for Single Crystals, NAS  Ukraine, Nauky Ave. 60, Kharkov 61001, Ukraine }

$^{2}$\textit{ V.N. Karazin Kharkiv National University 4 Svobody Sq., Kharkov 61022, Ukraine}

\bigskip

\begin{abstract}
In the present work, the behavior of vacancy pore inside of spherical particle is investigated. On the assumption of quasistationarity of diffusion fluxes, the nonlinear equation set was obtained analytically, that describes completely pore behavior inside of spherical particle. Limiting cases of small and large pores are considered. The comparison of numerical results with asymptotic behavior of considered limiting cases of small and large pores is discussed.
\end{abstract}


\section{Introduction}

In the present time, investigations are intensively developing of the properties of various meso- and nanosystems \cite{ABPMT09}-\cite{MR12}. Creation of meso- and nanosystems by various methods is, as a rule, accompanied by the formation of their defect structure. Properties of such particles to a large degree depend just on this defect structure \cite{YinEtAl04}-\cite{CRF08}. Such defects can be classified according to their dimensionality. For example, vacancies or interstitial atoms can be related to point or null-dimensional objects, while dislocations are referred to one-dimensional objects. Two-dimensional objects can be represented by grain boundaries of polycrystalline particles. The properties of such meso- and nanoparticles are especially strongly affected by three-dimensional defects. Vacancy pores, pores filled with gas and new phase inclusions are some of the most frequently encountered three-dimensional defects in such meso- and nanoprticles. Regularities of diffusion growth, motion and healing of such defects in nanoparticles preset an important problem. Such defect structure plays an important role in the further compactification of nanoparticles and creating new materials \cite{Rag07}. Establishing regularities of defect structure evolution will enable one to control it as well as to change properties of corresponding meso- and nanoparticles. The behavior of vancy pores in inorganic mediums is investigated in detail ( see e.g. \cite{Lif59}-\cite{DTY85}). The problem of pore growth in bounded particles is essentially more complicated. It is rather close to the problem of the interaction of pores in unbounded materials. The role of the second object, the pore is interacting with, is performed by particle boundary. The diffusion interaction of pores as well as of other new phase precipitations  in macroscopic materials has been investigated ( see e.g. \cite{Geg71}, \cite{Max76}, \cite{DTY85}), and the implemented methods are useful for solving the problem under discussion.

It is well known \cite{Geg71} , that the main factor causing swelling of solid bodies irradiated by neutrons is the formation of pores, that line up into periodic structure of lattice type. In order to explain this phenomenon, in work \cite{Max76} the theory was developed of diffusion interaction of pores. It has been shown in this work, on the example of two-pore interaction, that the larger pore of the radius $R_{B} $ ($R_{B} >R_{c} $, where $R_{c} $ is the critical radius of the pore) creates around itself lowered concentration of vacancies, thus retarding neighboring pore growth. On the contrary, the smaller pore of the radius $R_{S} $ ($R_{S} <R_{c} $) creates around itself, while evaporating, heightened vacancy concentration thus increasing growth rate of the neighboring pore. Heterogeneity of point defect distribution leads to motion of pores, that can lead to the situation when the smaller pore 'runs away' from the larger one, that, in turn, is chasing the smaller one. Such effects work well for distant pores. For close pores, the other effects manifest themselves. Thus, in the work \cite{DTY85}, the theory of diffusion interaction of pores at unconditioned distances has been created. It has been shown, as well, that, on close distances, the transition of vacancies from the smaller pore to the larger one is possible.

The creation of the theory of the diffusion interaction of pores in bounded medias, for example, in spherical nanoshells, is an exceptionally complicated task. In bounded matrices, the influence of a close boundary complicates strongly pore behavior and makes it fundamentally different from that in unbounded matrices.  The formation of pores in spherical nanoshells is a relatively recent discovery \cite{YinEtAl04}. In survey \cite{ZGP12} the results both of theoretical investigation and of computer modelling were presented, concerning formation and disappearing of pores in spherical and cylindrical nanoparticles. Great attention is devoted to the problem of stability of hollow nanoshells, i.e. of particles that with large vacancy pores in their centers \cite{FS08},\cite{GZTG05}.

In the present work the behavior of vacancy pore in solid-state spherical matrix is investigated. On the assumption of quasi equilibrium of diffusion fluxes, the canonical equations for evolution of pore radius, spherical matrix radius as well as center-to-center distance between the pore and the spherical matrix have been obtained analytically.  The obtained system of ordinary differential equations describes completely the evolution both of the pore and of the spherical matrix. The absence of critical pore size, unlike the case of an unbounded matrix, has been demonstrated. The extreme cases were considered when pore dynamics is simplified.  In general case, pores in such particles are dissolving diffusively while decreasing in size and moving towards the center of a spherical matrix. Main regularities of pore behavior inside a spherical particle have been established. The obtained results are of general character and are useful for comparison with regularities obtained as a  result of numerical modeling. It is just the obtained solutions, that should used for comparison with pore behavior, e.g. at numerical modeling. This will enable one  to find deviations from usual diffusive behavior of pores in small bounded particles.

\section{Equations for time evolution of a pore}

\begin{figure}
  \centering
  \includegraphics[width=6 cm]{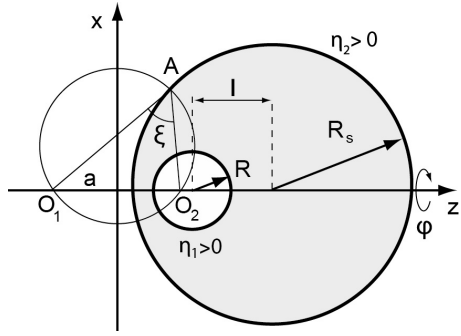}\\
  \caption{Pore in spherical granule in bispherical coordinate system. Pore and granule surfaces in this system are coordinate planes $\eta =\textrm{const}.$}  \label{fg1}
\end{figure}

Let us consider spherical granule of radius $R_s $ containing a vacancy pore of radius $R<R_s $ (see Fig. \ref{fg1} ). We denote values of these radii at the initial moment of time $t=0$ as $R_s $ and $R$ correspondingly. Let center-to-center distance between the pore and the granule be equal to $l$. We are interested in time evolution of the pore and the granule under the influence of diffusion fluxes of vacancies. The complete description of such evolution implies knowledge about the time change of pore and granule dimensions as well as of their center-to-center distance. In order to obtain the equations for evolution of these values, boundary conditions are necessary, that are determined by equilibrium concentration values near spherical surfaces of the pore and the granule. Equilibrium concentration of vacancies near the surface of a spherical pore in the absence of external pressure is determined by the relation (see e.g. \cite{ChSB90} ):
\begin{equation}\label{eq1}
    c_{R}=c_{V}\exp\left(\frac{2\gamma\omega}{kT
    R}\right)\,,
\end{equation}	
where $c_V $ is thermal-equilibrium vacancy concentration near plane surface, $\gamma $ is surface energy, $T$ is granule temperature, $\omega $  is volume related one lattice site.
In the same way, equilibrium concentration of vacancies near free surface of a spherical granule is determined (on assumption of the absence of external pressure). 	
	\begin{equation}\label{eq2}
   c_{R_s}=c_V\exp\left(-\frac{2\gamma\omega}{kTR_s}\right)\,,
\end{equation}	
These concentration values will determine vacancy fluxes. In the further consideration, we will suppose that equilibrium concentrations adjust quickly to the change of pore and granule sizes. In other words, equilibrium concentrations tune themselves to pore and granule size change. Certainly, the problem remains extremely complicated. For the sake of simplicity, it is natural to make one more assumption, namely, to suppose that stationary fluxes of vacancies inside granule are quickly established. There are two arguments in favor of this.
First of all, even if one gets out of the limits of such assumption, vacancy distribution inside granule is unknown. Besides, in a number of cases, stationary fluxes are established quickly enough. The evaluation of characteristic time of establishing stationary fluxes gives $\tau \ll l^2 /D $. Under such assumptions, diffusion flux of vacancies onto pore and granule boundaries is determined by stationary diffusion equation and corresponding boundary conditions
\begin{equation}\label{eq3}
  \Delta c =0 ,
\end{equation}
\[c(r)|_{r=R}=c_R ,\]
\[c(r)|_{r=R_s}=c_{R_s} .\]
The geometry of pore and granule boundaries dictates the use of bispherical coordinate system \cite{Arf70}, as the most convenient one. In bispherical coordinate system (see Fig. \ref{fg1}) each point $A$ of the space is matched to three numbers $(\eta ,\xi ,\varphi )$, where $\eta =\ln \left(\frac{|AO_{1} |}{|AO_{2} |} \right)$, $\xi =\angle O_{1} AO_{2} $, $\varphi $ is polar angle. Let us cite relations, that connect bispherical coordinates with Cartesian ones:
\begin{equation}\label{eq4}
  x=\frac{a\cdot \sin\xi\cdot\cos\varphi}{\cosh\eta-\cos\xi}, \quad
  y=\frac{a\cdot \sin\xi\cdot\sin\varphi}{\cosh\eta-\cos\xi}, \quad
  z=\frac{a\cdot \sinh\eta}{\cosh\eta-\cos\xi},
\end{equation}
 where $a$ is parameter, that at fixed values of pore and granule radii as well as of their center-to-center distance is determined by relation
 $$a=\frac{\sqrt{[(l-R)^2-R_s^2][(l+R)^2-R_s^2]}}{2\cdot l}\,.$$
Pore and granule surfaces in such coordinate system are given by relations 	
\begin{equation}\label{eq5}
    \eta_1=\textrm{arsinh} \left(\frac{a}{R}\right), \quad
\eta_2=\textrm{arsinh} \left(\frac{a}{R_s}\right).
\end{equation}
These relations determine values of $\eta_1 $ and $\eta_2 $ from pore and granule radii, while $a$ includes additionally center-to center distance between the pore and the granule. In the bispherical coordinate system the equation determining vacancy concentration and boundary condition takes on a following form:
\begin{equation}\label{eq6}
\frac{\partial}{\partial\eta}\left(\frac{1}{\cosh\eta-\cos\xi}\frac{\partial c}{\partial\eta}\right)+
  \frac{1}{\sin\xi}\frac{\partial}{\partial\xi}\left(\frac{\sin\xi}{\cosh\eta-
  \cos\xi}\frac{\partial c}{\partial\xi}\right)+\frac{1}{(\cosh\eta-
  \cos\xi)\cdot \sin^2\xi}\frac{\partial^2
  c}{\partial\varphi^2}= 0
\end{equation}
\[c(\eta , \xi, \varphi)|_{\eta_1}=c_R\]
\[c(\eta , \xi, \varphi)|_{\eta_2}=c_{R_s}\]
Due to symmetry of the problem, vacancy concentration does not depend on variable $\varphi $. Consequently, equation (\ref{eq6}) is reduced to
\begin{equation}\label{eq7}
    \frac{\partial}{\partial\eta}\left(\frac{1}{\cosh\eta-\cos\xi}\frac{\partial c}{\partial\eta}\right)+
  \frac{1}{\sin\xi}\frac{\partial}{\partial\xi}\left(\frac{\sin\xi}{\cosh\eta-
  \cos\xi}\frac{\partial c}{\partial\xi}\right)= 0
\end{equation}
Let us perform substitution for required function $c(\eta ,\xi )=\sqrt{\cosh \eta -\cos \xi } \cdot F(\eta ,\xi )$ that gives us equation for function $F(\eta ,\xi )$ in the following form:
\begin{equation}\label{eq8}
 \frac{\partial^2F}{\partial\eta^2}+\frac{1}{\sin\xi}\frac{\partial}{\partial\xi}
 \left(\sin\xi\frac{\partial F}{\partial\xi}\right)-\frac14F=0\,,
\end{equation}
Let us try solution of the equation by the method of separation of variables: $F(\eta ,\xi )=F_1 (\eta )\cdot F_2 (\xi )$. As a result, the following equations are obtained

\[\frac{d^2F_1}{d\eta^2}=\left(k+\frac{1}{2} \right)^2\cdot F_1 ,\]
\[ \frac{1}{\sin\xi}\frac{d}{d\xi}
 \left(\sin\xi\frac{d F_2}{d\xi}\right)=-k\cdot(k+1)\cdot F_2\,.\]
Here parameter $k$ is separation constant. The solution of these equations can be easily found, taking into account that the second one coincides with Legendre equation. Then, general solution can be written down in the form:
\begin{equation}\label{eq9}
c(\eta,\xi) = \sqrt{\cosh\eta-\cos\xi}\times$$
$$\times\sum_{k=1}^{\infty} (A_k\cdot\exp(k+1/2\eta)\cdot
P_k(\cos\xi)+B_k\cdot\exp(-(k+1/2\eta))\cdot P_k(\cos\xi))\,,
\end{equation}
where $A_k $ and $B_k $ are, for the time being, arbitrary constants, and $P_k (x)$ are Legendre polynoms

\[P_k(x) = \frac1{2^k\cdot k!}\frac{d^k}{dx^k}(x^2-1)^k,\,\quad P_0(x)\equiv 1.\]
We still have to determine the values of arbitrary constants from boundary conditions and find boundary problem solution (\ref{eq6}) as
\[c(\eta, \xi ) =
\sqrt{2(\cosh\eta-\cos\xi)}\left\{{c_R}\sum_{k=0}^\infty
\frac{\sinh(k+1/2)(\eta-\eta_2)}{\sinh(k+1/2)(\eta_1-\eta_2)}\exp(-(k+1/2)\eta_1)P_k(\cos\xi)-
\right.\]
\begin{equation}\label{eq10}
 \left.- c_{R_s}\sum_{k=0}^\infty
\frac{\sinh(k+1/2)(\eta-\eta_1)}{\sinh(k+1/2)(\eta_1-\eta_2)}\exp(-(k+1/2)\eta_2)P_k(\cos\xi)
\right\}\,.
\end{equation}
Let us note, that here boundary concentration $c_R $ is expressed through $\eta_1 $ and $a$, and $c_{R_s} $ through $\eta_2$ and $a$. This solution determines stationary vacancy concentration anywhere inside spherical granule of  radius $R_s $ and outside pore of radius $R$ (their center-to-center distance is equal to $l$). However, the knowledge of vacancy concentration allows one to find vacancy fluxes onto the pore as well as onto granule boundary at the given positions of granule and pore. These fluxes cause change of pore and granule sizes as well as of pore position. Thus, using the obtained values of the  fluxes, one can write down the equtions for the time change of pore and granule radii as well as of their center-to-center distance. Vacancy flux is determined by the first Fick's low as
\begin{equation}\label{eq11}
\vec{j}=-\frac {D}{\omega} \nabla c\,,
\end{equation}
where $D$ is diffusion coefficient. Let denote the outer pore surface normal as $\vec{n}$. Then vacancy flux onto pore surface is determined by scalar product $\vec{n} \cdot \vec{j}|_{\eta =\eta_1 } $. Let us write down the expression for vacancy flux onto unit area of pore surface using the expression for gradient in bispherical coordinates \cite{Arf70}
\begin{equation}\label{eq12}
\vec{n} \cdot \vec{j}|_{\eta=\eta_1}=\frac {D}{\omega} \cdot
\frac{\cosh\eta_1-\cos\xi}{a}\frac{\partial
c}{\partial\eta}\left|_{\eta=\eta_1}\right.\,.
\end{equation}
Similar expression determines vacancy flux onto unit area of granule surface
\begin{equation}\label{eq13}
\vec{n} \cdot \vec{j}|_{\eta=\eta_2}=\frac {D}{\omega} \cdot
\frac{\cosh\eta_2-\cos\xi}{a}\frac{\partial
c}{\partial\eta}\left|_{\eta=\eta_2}\right.\,.
\end{equation}
Here $\vec{n}$ is granule surface normal. Evidently, the total vacancy flux onto pore surface determines rate of change of pore volume. It is natural to suppose, that surface diffusion, whose diffusion coefficient usually much exceeds that of the bulk, is in time to restore spherical shape of the pore and the granule. Thus, it is easy to write down the equation for pore volume change in the form
\[\dot{R}=-\frac{\omega}{4\pi R^2} \oint\vec{n}\cdot\vec{j}|_{\eta=\eta_1} \, dS\]
In the same way one obtains the equation that determines granule radius:
\[\dot{R_s}=-\frac{\omega}{4\pi R_s^2}\oint \vec{n}\cdot\vec{j}|_{\eta=\eta_2} dS\]
After substitution of the exact solution and performing integration, one obtains an equation for pore radius change with time:
\begin{equation}\label{eq14}
    \dot{R}=-\frac{D}{R}\left[\frac{c_R}{2}
+\sinh\eta_1\cdot(c_R\cdot(\Phi_1+\Phi_2)-2c_{R_s}\cdot\Phi_2)\right],
\end{equation}
where functions $\Phi_1 $ and $\Phi_2 $ are introduced, that consist of the sum of exponential series:
\[\Phi_1=\sum_{k=0}^\infty \frac{ e^{-(2k+1)\eta_1}}{e^{(2k+1)(\eta_1-\eta_2)}-1}, \quad \Phi_2=\sum_{k=0}^\infty \frac{ e^{-(2k+1)\eta_2}}{e^{(2k+1)(\eta_1-\eta_2)}-1}\]
The details of the derivation are given in the appendix. Here $\eta_1 $ and $\eta_2 $ are expressed through pore and granule radii in correspondence with relations (\ref{eq5}) , while $c_R $ and $c_{R_s } $ are expressed through the same radii via relations (\ref{eq1}) and (\ref{eq2}). Thus, the right part of this equation depends nonlinearly on $R$, $R_s $ and $l$. In a similar way one obtains equation
\begin{equation}\label{eq15}
    \dot{R_s}=-\frac{D}{R_s}\left[\frac{c_{R_s}}{2}
+\sinh\eta_2\cdot(2c_{R}\cdot\Phi_2 - c_{R_s}\cdot(\Phi_2+\Phi_3)) \right]\,,
\end{equation}
where the following definition for the function $\Phi _{3} $ is introduced:
\[\Phi_3=\sum_{k=0}^\infty \frac{ e^{-(2k+1)(2\eta_2-\eta_1)}}{e^{(2k+1)(\eta_1-\eta_2)}-1}=\sum_{k=0}^\infty \frac{ e^{-(2k+1)\eta_3}}{e^{(2k+1)(\eta_1-\eta_2)}-1}\]
In order to obtain closed set of equations determining granule and pore evolution, one needs to complement these equations with one for the rate of changing center-to-center distance between the pore and the granule.  Of course, the displacement rate of vacancy pore relative to granule center is also determined by diffusion fluxes of vacancies onto pore surface (see e.g. \cite{Geg71,ChSB90}). In the present case,  the displacement rate is determined by relation
\begin{equation}\label{eq16}
\vec{v}=-\frac{3\omega}{4\pi R^2}\oint \vec{n}(\vec{n}\cdot\vec{j}_v)|_{\eta=\eta_1} dS.
\end{equation}
Using the exact solution (\ref{eq10}) and performing integration (see Appendix), one obtains:
\begin{equation}\label{eq17}
\vec{v}=\vec{e_z}\cdot\frac{3 D}{R}\times$$
$$\times
\left[\sinh^2\eta_1 \cdot(c_R\cdot
(\widetilde{\Phi}_1+\widetilde{\Phi}_2)-2c_{R_s}\cdot \widetilde{\Phi}_2)-\frac{1}{2}\sinh2\eta_1\cdot(c_R\cdot(\Phi_1+\Phi_2)-2c_{R_s}\cdot \Phi_2)\right]
\end{equation}
Here new functions $\widetilde{\Phi }_1 $ and $\widetilde{\Phi }_2 $ defined:
\[\widetilde{\Phi}_1=\sum_{k=0}^\infty \frac{(2k+1)e^{-(2k+1)\eta_1}}{e^{(2k+1)(\eta_1-\eta_2)}-1}, \quad \widetilde{\Phi}_2=\sum_{k=0}^\infty \frac{(2k+1)e^{-(2k+1)\eta_2}}{e^{(2k+1)(\eta_1-\eta_2)}-1}\,.\]
Taking into account that displacement rate along $z$ axis coincides with $dl/dt$, let us write down the equation in the final form
\begin{equation}\label{eq18}
   \frac{dl}{dt} =  \frac{3 D}{R}\times$$
$$\times
\left[\sinh^2\eta_1 \cdot(c_R\cdot
(\widetilde{\Phi}_1+\widetilde{\Phi}_2)-2c_{R_s}\cdot \widetilde{\Phi}_2)-\frac{1}{2}\sinh2\eta_1\cdot(c_R\cdot(\Phi_1+\Phi_2)-2c_{R_s}\cdot \Phi_2)\right]
\end{equation}
The obtained equation set (\ref{eq14}), (\ref{eq15}) and (\ref{eq18}) determines completely evolution of the pore and the granule with time. Let us discuss several general properties of the obtained equation set.   First of all, it is clear that the volume of granule material does not change with time. Vacancies only carry away 'emptiness'. It is easy to establish this conservation low from the obtained equation set. It can be shown easily that
\[R_s(t)^2\dot{R}_s(t)-R(t)^2\dot{R}(t)=0\]
The validity of such conservation low is connected closely with current quasistationary approximation. Vacancy  fluxes, that come out from the pore and from the granule are balanced with each other. Thus, the volumes of the pore and of the granule are connected with each other by an easy relation:
\begin{equation}\label{eq19}
  R_s(t)^3 ={V+R(t)^3}
\end{equation}
where $V=R_s(0)^3-R(0)^3 $ is initial volume of granule material (multiplier $4\pi /3$ is omitted for convenience). The existence of such conservation law allows diminishing a number of unknown quantities. It also can be proven, that expressions in the right parts of the given equations do not reverse signs. This circumstance points to monotonous diminishing with time of pore and granule radii as well as of center-to-center distance between the pore and the granule. An important conclusion can be made from the above said, that there exists no critical pore size. This is a cardinal difference from the pore evolution in unbounded materials. The validity of the conservation low allows one to investigate approximately system behavior, since the volume of the granule itself doesn't change essentially with time. Expression (\ref{eq19}) allows one to determine $R_s $ and reduce the problem to the set of two differential equations for $R$ and $l$. Below, some asymptotic regimes of the pore evolution inside the granule will be considered.

\section{Asymptotic evolution modes}

The obtained above nonlinear evolution equations are exceptionally complicated. Therefore, it makes sense to consider characteristic limiting cases when evolution equations take on tractable form. Such modes are determined by the relations of three dimensionless quantities  $R/R_s $, $l/R_s $ and $R/l$. These quantities are restricted geometrically by the following inequality:
\begin{equation}\label{eq20}
    R/R_s+l/R_s < 1
\end{equation}
This inequality means that the pore is situated inside the granule without touching its boundaries. If the surfaces of a pore and a granule touch each other, pore evolution will differ cardinally from its evolution due to vacancy fluxes. Such case should be considered separately with an account of healing sharp rims and is out of scope of the present work. Evidently, the value  $\delta =R/R_s < 1$ is always smaller then unity since pore size can't exceed that of the granule. It is clear from general physical considerations that, in the course of evolution, values $R/R_s $ and $l/R_s$ are diminishing. Let us discuss now what kinds of asymptotic modes can be realized.

\subsection{Small pores}

First of all, let us consider the case of small pores $R/R_s \ll 1$. Taking into account that size of the pore can only diminish, this approximation holds true over the entire pore evolution.  At this, distance from the pore to granule center can vary. Thus, the case is possible when
\[ R/R_s \ll 1, \quad l/R_s \ll 1,\]
At this, proportions of these values can vary. The possibility exists:
\[R/R_s \ll l/R_s  \Rightarrow R/l \ll 1\]
This means, that the distance from a small pore to the granule center is large as compared to granule radius.
	\begin{equation}\label{eq21}
 1) \quad R/R_s \ll 1, \quad l/R_s \ll 1,\quad R/l \ll 1
\end{equation}	
Of course, another disposition is possible, when a small pore is situated close to the granule center. In this case, the relation between the values is opposite:
\[R/R_s \gg l/R_s  \Rightarrow R/l \gg 1\]
Then, the next possible mode is determined by the relations of values
\begin{equation}\label{eq22}
   2) \quad R/R_s \ll 1, \quad l/R_s \ll 1,\quad R/l \gg 1
\end{equation}
Moreover, small pores can be situated at significant distance from the granule center that is comparable with granule size.  In this case, the next relation is realized:
\begin{equation}\label{eq23}
  3) \quad R/R_s \ll 1, \quad l/R_s \simeq 1, \quad R/l \ll 1.
\end{equation}
\begin{figure}
  \centering
  \includegraphics[ height=7 cm]{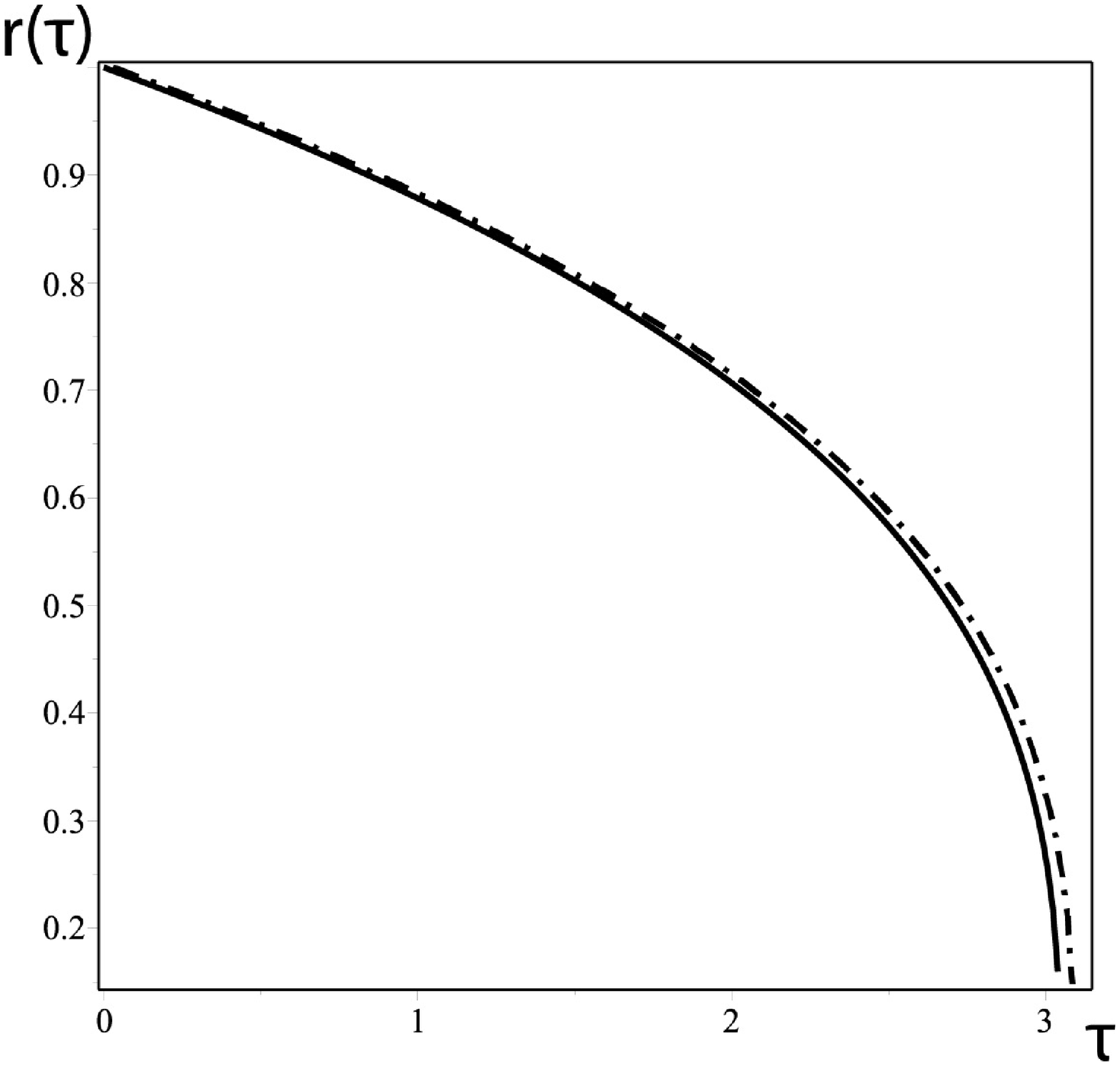}
	\includegraphics[ height=7 cm]{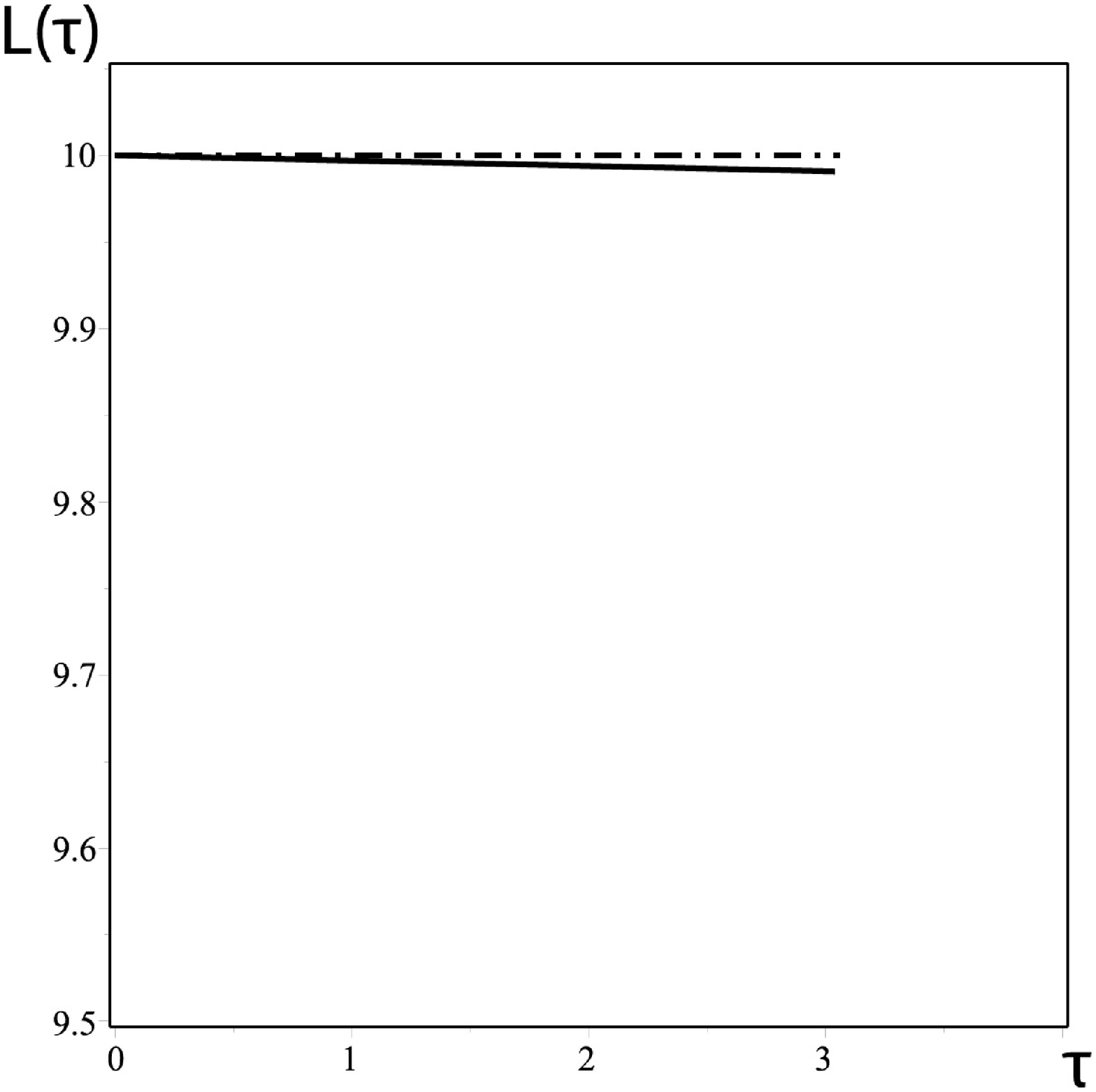}\\
  \caption{On the left, the plots are shown for time dependences of pore radius: solid line corresponds to numerical solution of complete equation set (\ref{eq25}), dash-and-dot line corresponds to the numerical solution of approximate equations (\ref{eq29})-(\ref{eq30}); On the right, solid line corresponds to numerical solution of complete equation set (\ref{eq25}), dash-and-dot line corresponds to the numerical solution of approximate equations (\ref{eq29})-(\ref{eq30}). The solutions are obtained at initial conditions $r|_{\tau =0} =1$, $r_{s} |_{\tau =0} =100$, $L|_{\tau =0} =10$ and $A=10^{-1} $.}
    \label{fg2}
\end{figure}
In this case, pore is situated close to granule boundary.

Let us note, that the case of small pores is distinguished by one more simplifying circumstance. It can be seen easily that healing of small pores $\frac{R(0)}{R_s(0)} \ll 1$ cannot be accompanied by a significant change of granule dimensions. Indeed, using the relation (\ref{eq19}), one can estimate an order of granule size change during the evolution. According to Eq. (\ref{eq19}) this change can be written down in the form:
\[\frac{R_s(t)}{R_s(0)}=\sqrt[3]{1-\frac{R(0)^3}{R_s(0)^3}+\frac{R(t)^3}{R_s(0)^3}} \simeq 1-\frac{1}{3}\frac{R(0)^3}{R_s(0)^3}  \]
Hence, within small pore approximation, granule size does not change $R_s(t) \approx R_s(0)=R_{s0}$ up to cubic order of smallness $\frac{R(0)^3}{R_s(0)^3} $. Then, neglecting granule radius change, we obtain Koshy problem for the set of two differential equations for $R$ and $l$, whose solution describes the evolution of the pore in the granules with time.

\begin{equation}\label{eq24}
\begin{cases}
  \frac{d l}{d t}=\frac{3Dc_V}{R}\cdot\exp\left(\frac{2\gamma\omega}{kTR}\right) \cdot \left[\frac{a^2}{R^2} \cdot (\widetilde \Phi _1  + \widetilde \Phi _2 ) - \frac{a}{R} \cdot \sqrt {1 + \frac{a^2}{R^2}}  \cdot (\Phi _1  + \Phi _2 )\right]-\\
-\frac{6Dc_V}{R}\cdot\exp\left(-\frac{2\gamma\omega}{kTR_{s0}}\right)\cdot \left[\frac{a^2}{R^2} \cdot \widetilde \Phi _2  - \frac{a}{R} \cdot \sqrt {1 + \frac{a^2}{R^2}}  \cdot \Phi _2 \right],\\
\frac{d R}{d t}=-\frac{Dc_V}{R}\cdot\exp\left(\frac{2\gamma\omega}{kTR}\right) \cdot \left[\frac{1}{2} + \frac{a}{R} \cdot (\Phi _1  + \Phi _2 )\right]+\frac{2Dc_V}{R}\cdot\exp\left(-\frac{2\gamma\omega}{kTR_{s0}}\right)\cdot \frac{a}{R} \cdot \Phi _2,\\
  R|_{t=0}=R(0),\\
  l|_{t=0}=l(0).
\end{cases}
\end{equation}

For the sake of convenience, let us make equation set (\ref{eq24}) dimensionless with a characteristic scale $R_0 =R(0)$ (that is pore radius at the initial time moment)  $t=0$ and characteristic time  $t_D =R_0^2/Dc_V $. Let us now go over to the following dimensionless variables:

\[r  = \frac{R}{R_0},\; r_s = \frac{R_{s}}{R_0},\;L = \frac{l}{R_0},\; {\tau} =\frac{t}{{t_D }},\;\alpha = \frac{a}{R_0},\;  \frac{{2\gamma \omega }}{{kTR}} = \frac{A}{{r}}, \;\frac{{2\gamma \omega }}{{kTR_{s0}}} = \frac{A}{{r_{s0}}},\; A = \frac{{2\gamma \omega }}{{kTR_0}}.\]

\begin{figure}
  \centering
 \includegraphics[width=6 cm]{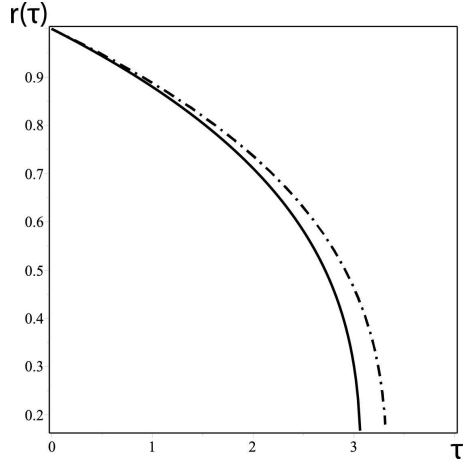}\\
\caption{Solid line represents numerical solution of the complete set of equations (\ref{eq25}), while dash-and-dot lite designates analytical solution of Eq. (\ref{eq31}) for initial conditions $r|_{\tau =0} =1$, $r_{s} |_{\tau =0} =100$, $L|_{\tau =0} =10$ and $A=10^{-1} $.}
\label{fg3}
\end{figure}
Then, the equation system (\ref{eq24}) can be rewritten in dimensionless form:
\begin{equation}\label{eq25}
\begin{cases}
  \frac{d L}{d \tau}=\frac{3\exp\left(\frac{A}{r}\right)}{r} \cdot \left[\frac{\alpha^2}{r^2} \cdot (\widetilde \Phi _1  + \widetilde \Phi _2 ) - \frac{\alpha}{r} \cdot \sqrt {1 + \frac{\alpha^2}{r^2}}  \cdot (\Phi _1  + \Phi _2 )\right]-\\
-\frac{6\exp\left(-\frac{A}{r_{s0}}\right)}{r}\cdot \left[\frac{\alpha^2}{r^2} \cdot \widetilde \Phi _2  - \frac{\alpha}{r} \cdot \sqrt {1 + \frac{\alpha^2}{r^2}}  \cdot \Phi _2 \right],\\
\frac{d r}{d \tau}=-\frac{\exp\left(\frac{A}{r}\right)}{r} \cdot \left[\frac{1}{2} + \frac{\alpha}{r} \cdot (\Phi _1  + \Phi _2 )\right]+\frac{2\exp\left(-\frac{A}{r_{s0}}\right)}{r}\cdot \frac{\alpha}{r} \cdot \Phi _2,\\
  r|_{\tau =0}=1,\\
  L|_{\tau =0}=\frac{l(0)}{R(0)}.
\end{cases}
\end{equation}
Let us now consider asymptotic case (\ref{eq21}). Let us present parameter $\alpha$ in Eqs. (\ref{eq25}) in the following form:
\[\alpha = \frac{r_{s0}^2}{2L}\sqrt{1+\left(\frac{L^2}{r_{s0}^2}-\frac{r^2}{r_{s0}^2}\right)^2-2\left(\frac{L^2}{r_{s0}^2}+\frac{r^2}{r_{s0}^2}\right)}.\]
Since $L \gg r$, the expression for parameter $\alpha $ is simplified
\begin{equation}\label{eq26}
\alpha \approx \frac{r_{s0}^2}{2L}\sqrt{\left(1-\frac{L^2}{r_{s0}^2}\right)^2}= \frac{r_{s0}^2}{{2L}}\left(1-\frac{L^2}{r_{s0}^2}\right),\end{equation}
 while bispherical coordinates $\eta _{1,2} $, determined by Eq.(\ref{eq5}) , are correspondingly, equal to:
\begin{equation}\label{eq27}
 \eta _1  = \textrm{arsinh} \left( {\frac{r_{s0}^2}{{2rL}}}\left(1-\frac{L^2}{r_{s0}^2}\right) \right),\;\eta _2  = \textrm{arsinh} \left( \frac{r_{s0}}{2L} \left(1-\frac{L^2}{r_{s0}^2}\right) \right).\end{equation}
Since $\frac{\sinh \eta_1 }{\sinh \eta_2 } =\frac{r_{s0}}{r}  \gg 1$, then $\eta_1 \gg \eta_2 $. In this case, sums of series can be estimated via following expressions:
\begin{equation}\label{eq28}
 \Phi_1 \approx \frac{1}{2\sinh2\eta_1},\; \Phi_2 \approx \frac{1}{2\sinh\eta_1},\;  \widetilde{\Phi}_1 \approx \frac{1+2\sinh^2\eta_1}{8\sinh^2\eta_1\cosh^2\eta_1},\; \widetilde{\Phi}_2 \approx \frac{\cosh\eta_1}{2\sinh^2\eta_1},$$
 $$ \sinh\eta_1=\frac{\alpha}{r},\quad \cosh\eta_1=\sqrt{1+\frac{\alpha^2}{r^2}}. \end{equation}

\begin{figure}
  \centering
 \includegraphics[ height=6 cm]{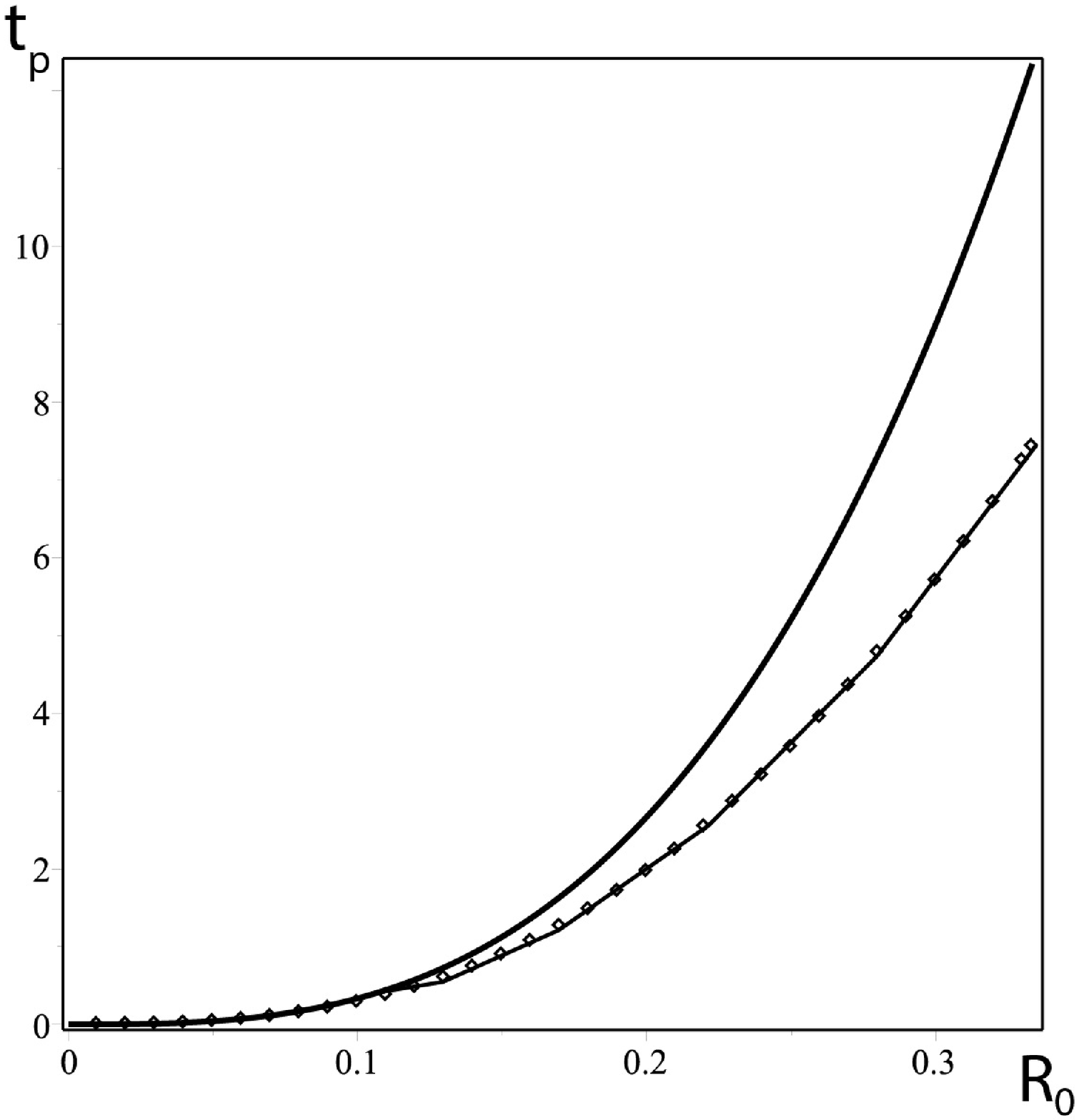}
\includegraphics[ height=6 cm]{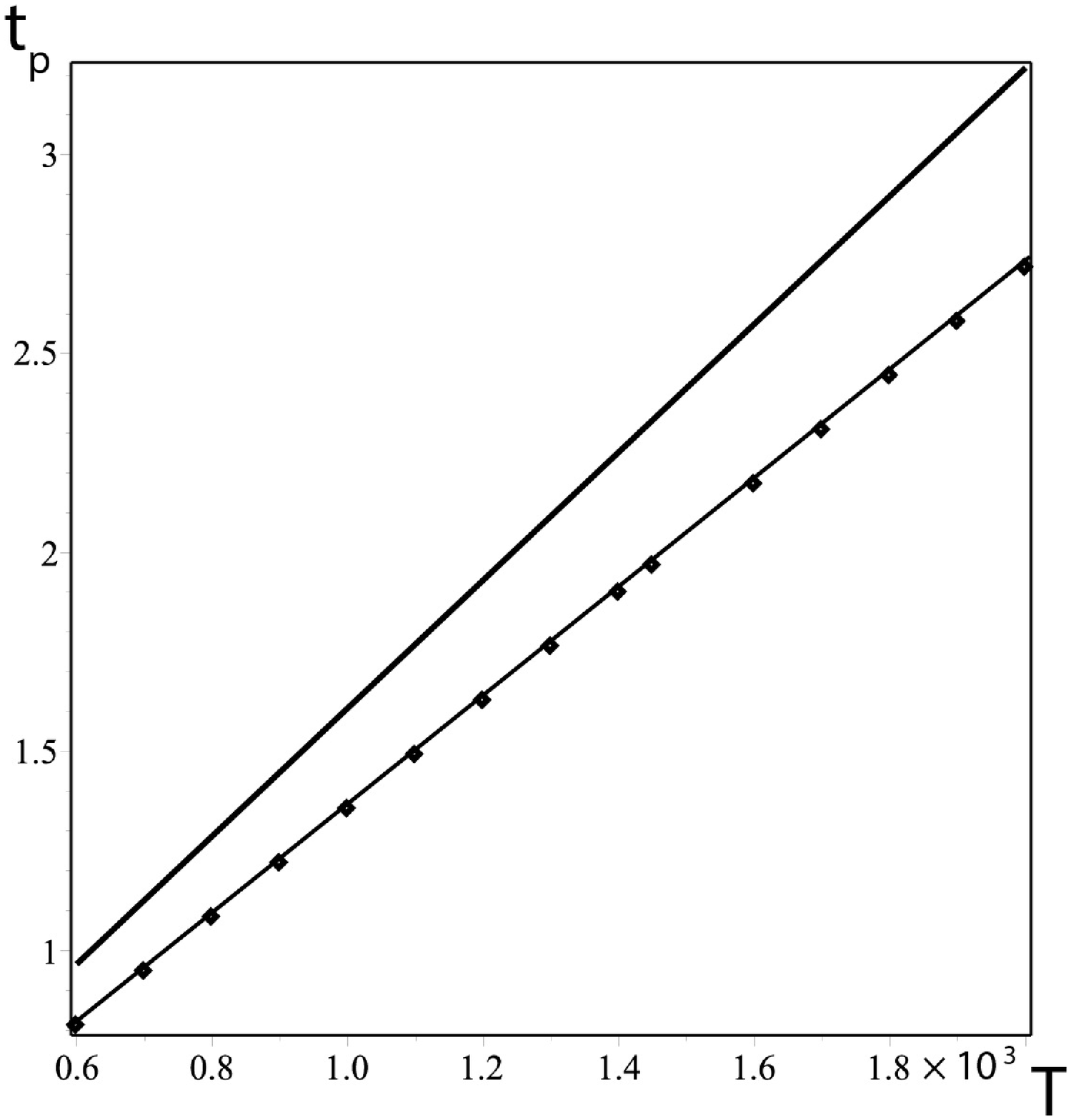}\\
\caption{In the left plot the dependence is shown of healing time $t_p $ on the change of relative pore radius $R_0=R(0)/R_s(0)$ for fixed values $R_s=10^{-4} $ cm, $l=10^{-5} $ cm and material (granule) temperature $T=1450^\circ \textrm{K}$; the right plot demonstrates dependence of $t_p $ on material (granule temperature) over the range $T\in \left[600^\circ \textrm{K},2000^\circ \textrm{K} \right]$  for fixed pore characteristics $R(0)=2\cdot 10^{-5}\textrm{cm}$ and $l_0= 10^{-5}\textrm{cm}$.}
\label{fg4}
\end{figure}

By substituting expressions (\ref{eq26})-(\ref{eq28}) into equation set (\ref{eq25}),  we find simplified equation set
\begin{equation}\label{eq29}
\frac{d L}{d \tau}=-\frac{3}{2}\,\exp\left(\frac{A}{r}\right)\cdot\frac{r \left(\frac{L^2}{r_{s0}^2}\right)}{r_{s0}^2} ,\end{equation}
\begin{equation}\label{eq30}	
\frac{d r}{d \tau}=-\frac{\exp\left(\frac{A}{r}\right)}{r} \cdot \left[1 + \frac{r}{2L}\cdot\left(\frac{L^2}{r_{s0}^2}\right) \right]+\frac{\exp\left(-\frac{A}{r_{s0}}\right)}{r}
\end{equation}
 Equations (\ref{eq29})-(\ref{eq30}) are written down up to $L^2 /r_{s0}^{2} $ terms.  This nonlinear set signifies that pore size diminishes monotonously while moving towards the granule center. Besides, taking into account smallness of the right part of Eq. (\ref{eq29}), it is easy to understand that pore motion towards granule center is slow. With account of finite time of pore healing, this means that displacement of the pore towards the granule center is insignificant.

It is interesting to compare the behavior of the pore in this asymptotic mode with the solutions of complete equation set (\ref{eq25}). In the Fig. \ref{fg2} , the numerical solutions of exact  (\ref{eq25}) and approximate   (\ref{eq29})-(\ref{eq30}) equation sets are shown with the same initial conditions $r|_{\tau =0} =1$, $r_{s} |_{\tau =0} =100$, $L|_{\tau =0} =10$ and $A=10^{-1} $. The left part of Fig. \ref{fg2}  demonstrates good agreement of the approximate solution with the solution of the complete equation set for pore radius time change.  In the right part of Fig. \ref{fg2} , the plot is shown for the time change of center-to-center distance between the pore and the granule. Over the time of pore healing, pore displacement towards the granule center is small in both cases. In the case of approximate solution, pore displacement rate is somewhat underestimated. Such good agreement allows us to consider pore radius change at zeroth-order  at $L^2 /r_{s0}^{2}  \ll 1$. In this case we obtain a simple equation for the radius of an immobile pore:
\begin{equation}\label{eq31}
\frac{d r}{d \tau}=-\frac{\exp\left(\frac{A}{r}\right)}{r}+\frac{\exp\left(-\frac{A}{r_{s0}}\right)}{r}\end{equation}
 It is easy to prove that the right part of this equation is negative. Therefore, the pore can only diminish in size. At this, two characteristic evolution stages exist. While $A \ll r$, the rate
\begin{figure}
  \centering
  \includegraphics[height=7 cm]{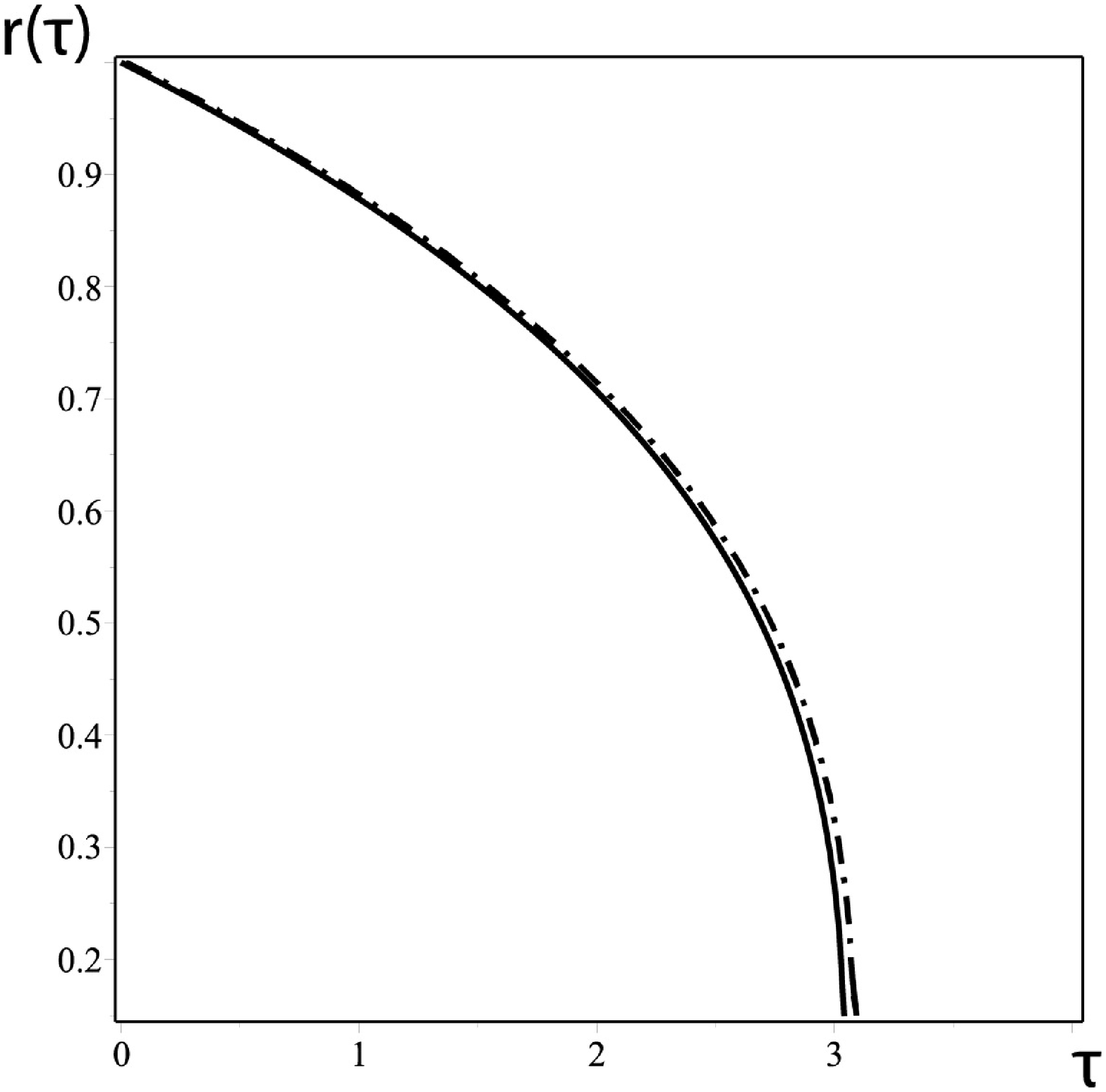}
	\includegraphics[ height=7 cm]{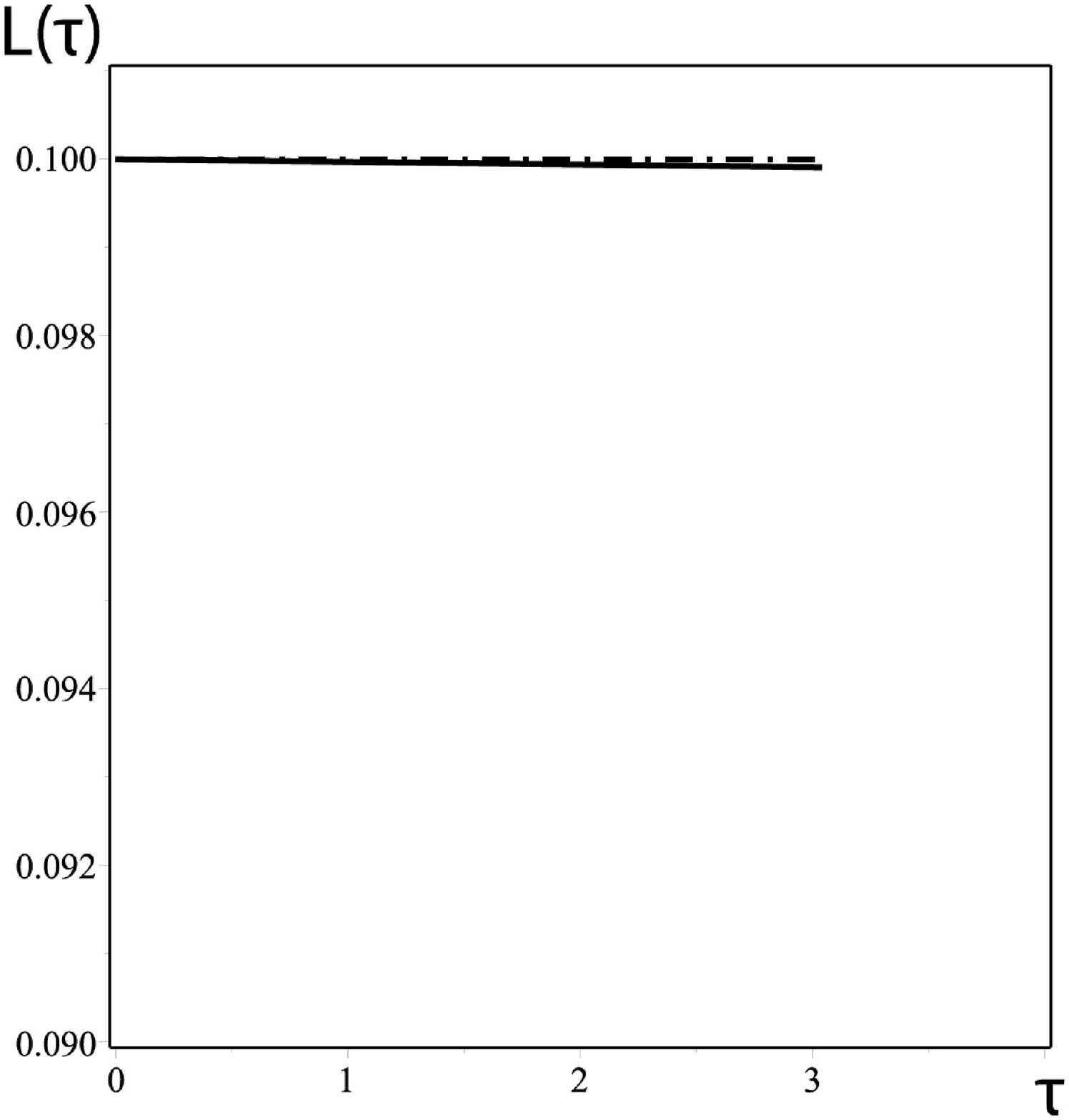}\\
\caption{On the left the plots are shown for the dependence of pore radius on time: solid line corresponds to numerical solution of equation set (\ref{eq25}), dash-and-dot line corresponds to numerical solution of equation set (\ref{eq40})-(\ref{eq41}); on the right, the plot is shown for the time change of center-to-center distance between the pore and the granule, obtained via numerical solution of Eq. (\ref{eq25}), dash-and-dot line indicates the  numerical solution of Eqs. (\ref{eq40})-(\ref{eq41}). All solutions are obtained under initial conditions $r|_{\tau =0} =1$, $r_{s} |_{\tau =0} =100$, $L|_{\tau =0} =0.1$ and $A=10^{-1} $.}
\label{fg5}
\end{figure}	
of size diminishing is small enough and increases with the decrease of radius as  $\sim r^{-2}$. The second stage ensues after reaching $A \sim r$ when pore dissolving rate starts to grow exponentially, causing 'instant' dissolving of the pore.  One can easily obtain solution of Eq. (\ref{eq31})  in integral form
\begin{equation}\label{eq32} \tau+\textrm{const}= e^{A/r_{s0}}\int\frac{rdr}{1-e^{A/r+A/r_{s0}}} \end{equation}
Supposing that most prolonged stage of pore dissolving occurs under conditions $A \ll r$ and $r \ll r_{s0} $, it is easy to conduct integration in (\ref{eq29}):
\begin{equation}\label{eq33} \tau \approx -\frac{\exp\left(\frac{A}{r_{s0}}\right)}{A}\left(\frac{r(t)^3}{3}-\frac{r(0)^3}{3}\right)\end{equation}
Here we neglected the fast second stage. In Fig. \ref{fg3}, the dashed line designates the plot of analytical solution (\ref{eq33}), that with a very small inaccuracy agrees with numerical solution of equation system (\ref{eq23}). Thus, at zeroth approximation, the healing of almost motionless pore occurs.

One of the most important characteristics of pore evolution is pore healing time. Taking into account the successful estimate of pore size change, we can estimate, using Eq. (\ref{eq33}), the characteristic healing time:

\begin{equation}\label{eq34} {\tau}_p \approx \frac{r(0)^3}{3A},\quad  \textrm{for} \quad A \ll r(0). \end{equation}

As it can be seen from Eq. (\ref{eq34}) , the healing time $\tau _p $ is changing by a cubic law, as a function of initial pore radius $r_0$, and by a linear law, as a function of material temperature $T$, because $A \sim 1/T$. The relation (\ref{eq34}) can be obtained from simple physical considerations. Indeed, vacancy flux $\vec{J}$ carries pore volume $V=\frac{4\pi}{3} R(0)^3$ through granule surface with an area  $S=4\pi R_s^2$  during healing time $t_{p} $:
\[R_s^2|\vec J|t_p \approx \frac{R(0)^3}{3}.\]
Estimating the flux as $|\vec J| \approx \frac{D c_V}{R(0)}\left(e^{A/R(0)}-e^{-{A/R_s}}\right)$ , one finds the relation for healing time
\[t_p \approx \frac{R(0)^4}{3R_s^2Dc_V \left(e^{A/R(0)}-e^{-{A/R_s}}\right)}.\]

Within the range $A/R(0) \ll 1$ and $A/R_s \ll 1$, we obtain $t_p \approx R(0)^5/3AR_s^2 Dc_V$ or, in dimensionless units $\tau _p  \simeq r(0)^{3} /3A $, that coincides with formula (\ref{eq32}).

On the left side of Fig. \ref{fg4}, the plot is shown for the dependence of pore healing time $\tau_p $ on initial relative pore radius $R_0 =R(0)/R_s(0)$. The solid line corresponds to formula (\ref{eq34}) while dots correspond to $t_p $ values, obtained by solving numerically  equation set (\ref{eq25}). One can observe the good agreement of the analytical result with the numerical one in the range of small ratios $R(0)/R_s(0)$, or small pores. On the right side of Fig. \ref{fg4}, the temperature dependences of pore healing time are shown, obtained by solving equation set (\ref{eq25}) both analytically and numerically. It is easy to note the linearity of the both temperature dependences. Thus, relation (\ref{eq33}) predicts well the healing time of small pores. In the plotting the time-dependence of pore healing time, the temperature dependence of diffusion coefficient was not taken into account (for solid matter, where $D=D_0 e^{-\frac{U}{kT}}$, where $U$ is vacancy activation energy. Naturally, if this dependence is taken into account, pore life time will decrease with temperature.  However, the relation  is valid also for a liquid phase, where temperature dependence of diffusion coefficient can be different. Thus, for generality,   temperature dependence of diffusion coefficient was not used. The plotted is only the temperature dependence, that is explicitly included into boundary conditions.

\begin{figure}
  \centering
 \includegraphics[width=6 cm]{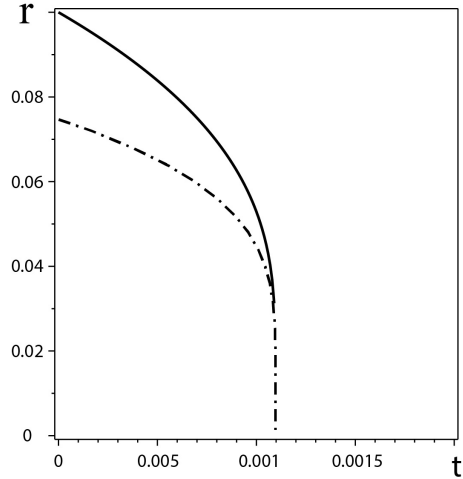}\\
\caption{Explosive mode of pore healing is demonstrated. Solid line designates the numerical solution of Eq.(\ref{eq35}), while dashed line shows the analytical solution (\ref{eq35}). Corresponding parameters are  $r|_{\tau =0} =0.1$, $r_{s} |_{\tau =0} =100$, $L|_{\tau =0} =10$ and $A=10^{-1} $. }
\label{fg6}
\end{figure}

Let us now consider in more details the fast, explosive stage of pore evolution. Such mode is realized after reaching by pore radius $r$ values of the order of $A$ ($r\approx A$ as it was noted before). After that, 'instant' pore dissolving follows. In order to analyze such a mode, one can neglect $\exp \left(-A/r_{s0} \right)$ in Eq. (\ref{eq31}). Then, the equation for pore radius evolution takes up the following form:
\begin{equation}\label{eq35} \frac{d r}{d \tau}=-\frac{\exp\left(\frac{A}{r}\right)}{r} \end{equation}
 The solution of Eq.(\ref{eq35}) can be expressed  through exponential integral
\begin{equation}\label{eq36} \tau_p-\tau=r^2\cdot \textrm{E}_3\left(\frac{A}{r}\right), \end{equation}
where $\textrm{E}_m(z)=\int_z^{\infty} e^{-x}\frac{dx}{x^{m}} $  is  exponential integral, $m$ is an integer number. When solving this equation, the arbitrary constant was chosen from the condition that, at the moment $\tau_p $, the pore disappears.
\begin{figure}
  \centering
 \includegraphics[width=6 cm]{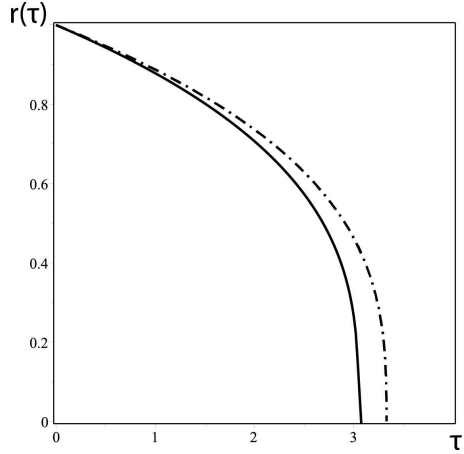}\\
\caption{Solid line demonstrates the numerical solution of equation set (\ref{eq25}), while dash-and-dot line shows the analytical solution of Eq. (\ref{eq41}) under the condition $L(\tau )\approx L(0)$ for initial conditions $r|_{\tau =0} =1$, $r_{s} |_{\tau =0} =100$, $L|_{\tau =0} =0.1$ and $A=10^{-1} $. }
\label{fg7}
\end{figure}
Since the right part of Eq. (\ref{eq35}) grows fast, we need to find an asymptotic solution at $r \ll A$. In the leading order, asymptotic of the exponential integral equals to: $\textrm{E}_3(A/r)\approx \frac{r}{A}\cdot e^{-A/r}$.  Using this asymptotic expansion, we find from formula (\ref{eq36}) time dependence of pore radius:
\begin{equation}\label{eq37} r(\tau)\approx -\frac{A}{\ln\left({A}({\tau_p-\tau})\right)} \end{equation}
It can be seen from this relation, that, during the finite time $\tau_p $, the pore dissolves completely. The explosive mode of pore healing is shown in Fig. \ref{fg6}. Thus, such mode always comes as a final stage of pore healing.

Let us now turn to the case (\ref{eq22}) of a small pore situated close to the granule center:

\begin{equation}\label{eq38}
 R/R_s \ll 1,\quad l/R_s \ll 1,\quad  R \gg l. \end{equation}

Such inequalities comply with the geometrical condition $R/R_s +l/R_s \leq 1$. Taking into account (\ref{eq38}), it is easy to find expressions for parameter $\alpha $ and coordinates $\eta _{1,2} $:
\begin{equation}\label{eq39} \alpha \approx \frac{r_{s0}^2}{2L}\left(1- \frac{r^2}{r_{s0}^2}\right),\; \eta _1  \approx \textrm{arsinh} \left( {\frac{r_{s0}^2}{{2rL}}}\left(1-\frac{r^2}{r_{s0}^2}\right) \right),\;\eta _2  \approx \textrm{arsinh} \left( \frac{r_{s0}}{2L} \left(1-\frac{r^2}{r_{s0}^2}\right) \right)
\end{equation}
It can be seen from here, that, for small pores, the relation $\eta_1 \gg \eta_2 $ is valid. Using the estimation of series sums, we obtain, from formulas (\ref{eq28}), approximate pore evolution equations for such case.
\begin{equation}\label{eq40} \frac{d L}{d \tau}=-\frac{3}{2}\cdot \exp\left(\frac{A}{r}\right)\cdot \frac{r\left(\frac{L}{r_{s0}}\right)^2}{r_{s0}^2 \left(1-\frac{r^2}{r_{s0}^2}\right)^2} \end{equation}
\begin{equation}\label{eq41} \frac{d r}{d \tau}=-\frac{\exp\left(\frac{A}{r}\right)}{r}\cdot \left[1+\frac{1}{2}\cdot \frac{rL}{r_{s0}^2\left(1-\frac{r^2}{r_{s0}^2}\right)}\right]+\frac{\exp\left(-\frac{A}{r_{s0}}\right)}{r} \end{equation}
In the Fig. \ref{fg5}, numerical solutions of Eqs. (\ref{eq25}) (solid line)  and Eqs. (\ref{eq40})-(\ref{eq41}) (dashed line) are shown for initial conditions, satisfying inequalities (\ref{eq38}): $r|_{\tau =0} =1$, $r_{s} |_{\tau =0} =100$, $L|_{\tau =0} =0.1$ and $A=10^{-1} $. The left part of Fig. \ref{fg5}   demonstrates a good agreement of numerical solutions
\begin{figure}
  \centering
  \includegraphics[height=7 cm]{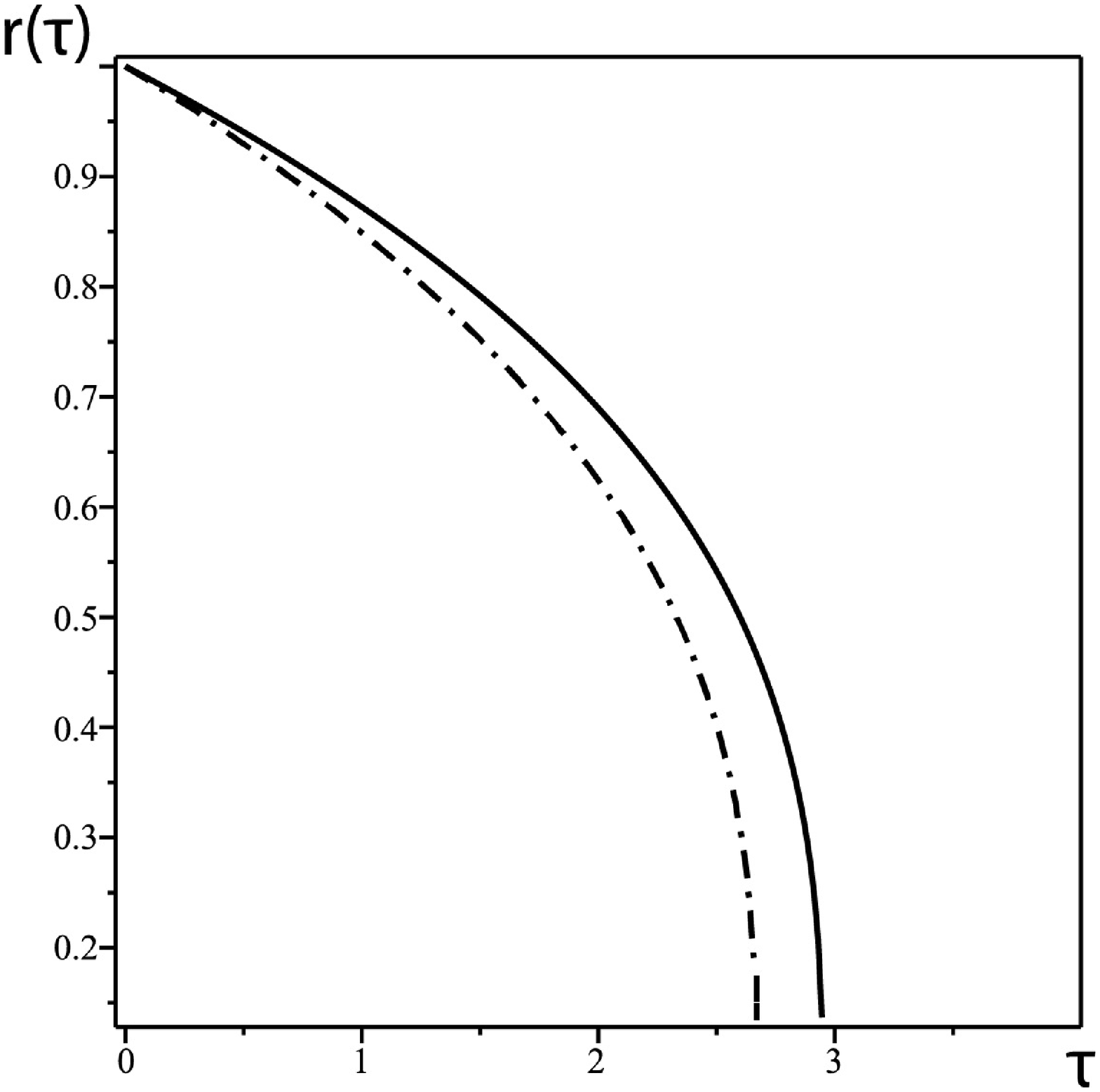}
	\includegraphics[height=7 cm]{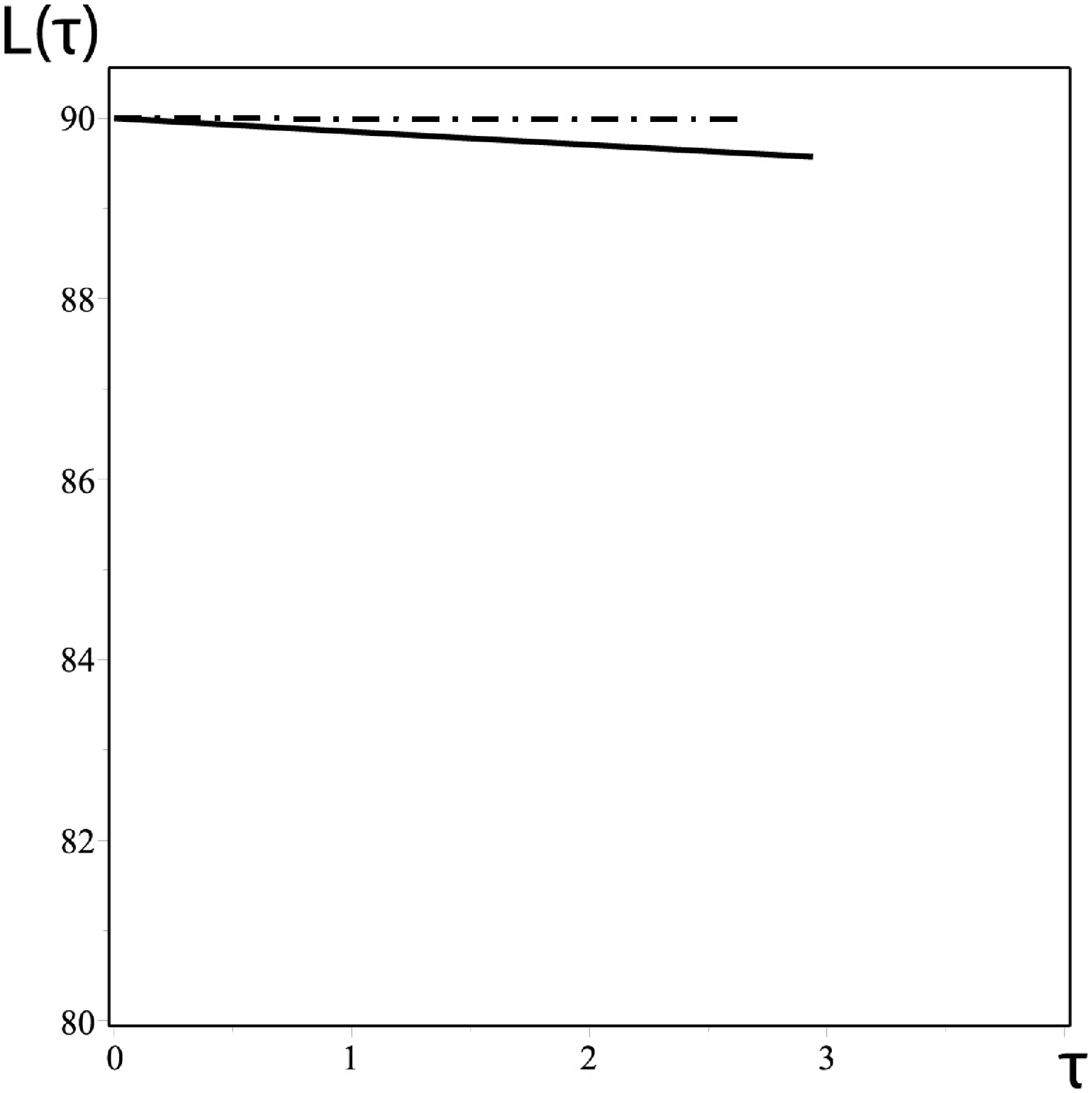}\\
	\caption{On the left, the plots are shown of the pore radius time dependence: solid line designates the numerical solution of equation set (\ref{eq40a})-(\ref{eq41a}) ; on the right, time dependence for distance $L(\tau )$ is presented: solid line designates the numerical solution of Eqs. (\ref{eq25}) , dash-and-dot line relates to the numerical solution of Eqs. (\ref{eq40a})-(\ref{eq41a}). All solutions are obtained for initial conditions $r|_{\tau =0} =1$, $r_{s} |_{\tau =0} =100$, $L|_{\tau =0} =90$ and $A=10^{-1} $.}
\label{fg8}
\end{figure}
of Eqs. (\ref{eq25})  and (\ref{eq40})-(\ref{eq41}) for pore radius change. In the right part of Fig. \ref{fg5}, time change of center-to-center distance between the pore and the granule is demonstrated for equation set (\ref{eq25}) and Eqs. (\ref{eq40})-(\ref{eq41}). Similarly to the previous case, the pore is almost immobile: $L(t)\approx L(0)$. Therefore, we can confine ourselves to zeroth approximation for the pore evolution analysis. In this case, Eq. (\ref{eq31}) is obtained. The analytical solution of this equation well agrees with the numerical solution of equation set (\ref{eq25}) for initial conditions $r|_{\tau =0} =1$, $r_{s} |_{\tau =0} =100$, $L|_{\tau =0} =0.1$ and $A=10^{-1} $. These solutions are shown in Fig. \ref{fg7}. The estimate of characteristic pore healing time coincides with formula  (\ref{eq34}).

\begin{figure}
  \centering
 \includegraphics[width=6 cm]{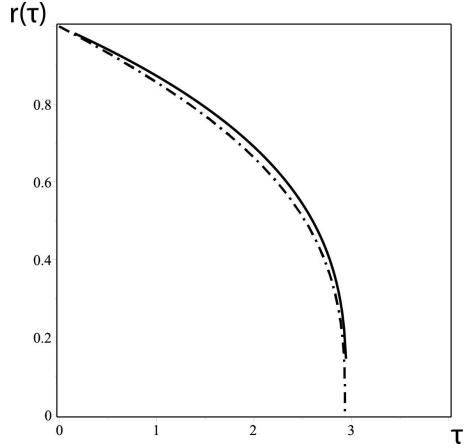}\\
\caption{In the figure, solid line represents numerical solution of equation set (\ref{eq25}), while dash-and-dot line relates to the analytical solution of Eq. (\ref{eq41a}) under condition $L(\tau )\approx L(0)$ for initial data $r|_{\tau =0} =1$, $r_{s} |_{\tau =0} =100$, $L|_{\tau =0} =90$ and $A=10^{-1} $.}
\label{fg9}
\end{figure}

Let us, finally, turn to the discussion of the mode (\ref{eq23}), when the pore is situated close to the granule boundary. In this case, the relation $l/R_{s} $ is close to unity:
\[\frac{l}{R_{s} } =1-\varepsilon ,\]
Here $\varepsilon $ is the small parameter, on which the asymptotic expansion is conducted. Parameter $\varepsilon $ value is restricted by the geometrical inequality (the pore inside the granule)
\[\frac{R}{R_{s} } \le \varepsilon .\]
In asymptotic expansion we will take into account the terms of the order of $\varepsilon^2 $. With account of this remark, parameter $\alpha $, and, correspondingly, bispherical coordinates $\eta_{1,2} $, obtained within the small pore approximation $R \ll R_s $ and $R \ll l$, take on the form:
\begin{equation}\label{eq42}
 \alpha \approx \frac{r_{s0}^2}{2L}\varepsilon(2-\varepsilon), \; \eta_1 \approx \ln \left(\frac{r_{s0}}{r}(2\varepsilon+\varepsilon^2)\right),\; \eta_2 \approx \varepsilon + \frac{\varepsilon^2}{2}. \end{equation}
Hence, $\eta_1  \gg \eta_2 $, therefore we can use previous estimates for the sums of series given by formulas (\ref{eq28}). Substituting (\ref{eq28}) and (\ref{eq42}) into the right part of Eq. (\ref{eq25}), we obtain pore evolution equations within approximation (\ref{eq23}):
\begin{equation}\label{eq40a} \frac{d L}{d \tau}=-\frac{3}{8}\cdot \exp\left(\frac{A}{r}\right)\cdot \frac{r\left(\frac{L}{r_{s0}}\right)^2}{r_{s0}^2 \left(1-\frac{L}{r_{s0}}\right)^2} \end{equation}
\begin{equation}\label{eq41a} \frac{d r}{d \tau}=-\frac{\exp\left(\frac{A}{r}\right)}{r}\cdot \left[1+\frac{1}{2}\cdot \frac{rL}{r_{s0}^2\left(1-\frac{L^2}{r_{s0}^2}\right)}\right]+\frac{\exp\left(-\frac{A}{r_{s0}}\right)}{r} \end{equation}
In Fig. \ref{fg8}, the numerical solutions are shown both of the exact equation set (\ref{eq25}) and of the approximate one (\ref{eq40a})-(\ref{eq41a}) with the same initial conditions $r|_{\tau =0} =1$, $r_{s} |_{\tau =0} =100$, $L|_{\tau =0} =90$ and $A=10^{-1} $. The left part of Fig. \ref{fg8} demonstrates very good agreement of the time dependences of pore radius. On the right part of Fig. \ref{fg8}, the plots are shown for time dependence of the center-to-center distance between the pore and the granule. It can be seen from the figure, that the displacement of the pore towards the granule center, obtained from exact equation set (\ref{eq25}), exceeds that observed in  approximate equation set (\ref{eq40a})-(\ref{eq41a}). However, in both cases, the displacement of the pore is small. However, in both cases, the displacement over the time of pore dissolving is small. If we confine ourselves to zero approximation $L(\tau )\approx L(0)$, the set (\ref{eq40a})-(\ref{eq41a}) is decoupled, which allows us to solve the equation for pore radius evolution (\ref{eq41a}).  Within the range $A/r \ll 1$ and $A/r_{s0}  \ll 1$, integration (\ref{eq41a}) gives the following result:
\begin{equation}\label{eq43}
 \tau =-\frac{\exp\left(\frac{A}{r_{s0}}\right)}{A}\left(\frac{r(t)^3}{3}-\frac{r(t)^6}{\beta}-\frac{r(0)^3}{3}+\frac{r(0)^6}{\beta}\right),
\end{equation}
where $\beta=12Ar_{s0}\left(3-\frac{4L(0)}{r_{s0}}+\frac{L(0)^2}{r_{s0}^2}\right).$
In Fig. \ref{fg9},  dashed line indicates the plot of analytical solution of Eq. (\ref{eq41a}) for the almost motionless pore, that well agrees with the numerical solution of equation set (\ref{eq25}). It is easy to find the characteristic pore healing time from the expression (\ref{eq43}):
\begin{equation}\label{eq44}
  {\tau}_p \approx \frac{r(0)^3}{3A}\left(1-\frac{3r(0)^3}{\beta}\right),\quad \textrm{for} \quad A \ll r_{s0}.
\end{equation}
Thus, neglecting the finite time of explosive pore healing does not essentially affect pore healing time value.

\subsection{Large pores}

Let us now proceed to discussing the evolution of large pores. Let us begin with the notion, that asymptotic mode
\begin{equation}\label{eq45}
 R/R_s \cong 1,\quad l/R_s \cong 1,\quad  R/l \cong 1. \end{equation}
is not, in fact, realized. Indeed, let us take into account the closeness of the two firs relations to the unity
\begin{equation}\label{eq46}
\frac{R}{R_s}=1-\varepsilon_1,\quad \frac{l}{R_s}=1-\varepsilon_2, \end{equation}
 where $\varepsilon_1  \ll 1$ and $\varepsilon_2 \ll 1$ are small parameters. Substituting (\ref{eq46}) into geometrical condition (\ref{eq20}), one finds: $1 \leq \varepsilon_1+\varepsilon_2$. Since $\varepsilon_{1,2} $ are small parameters, this inequality does not hold. Thus, mode (\ref{eq45}) is not compatible with geometrical condition (\ref{eq20}).

\begin{figure}
  \centering
  \includegraphics[height=7 cm]{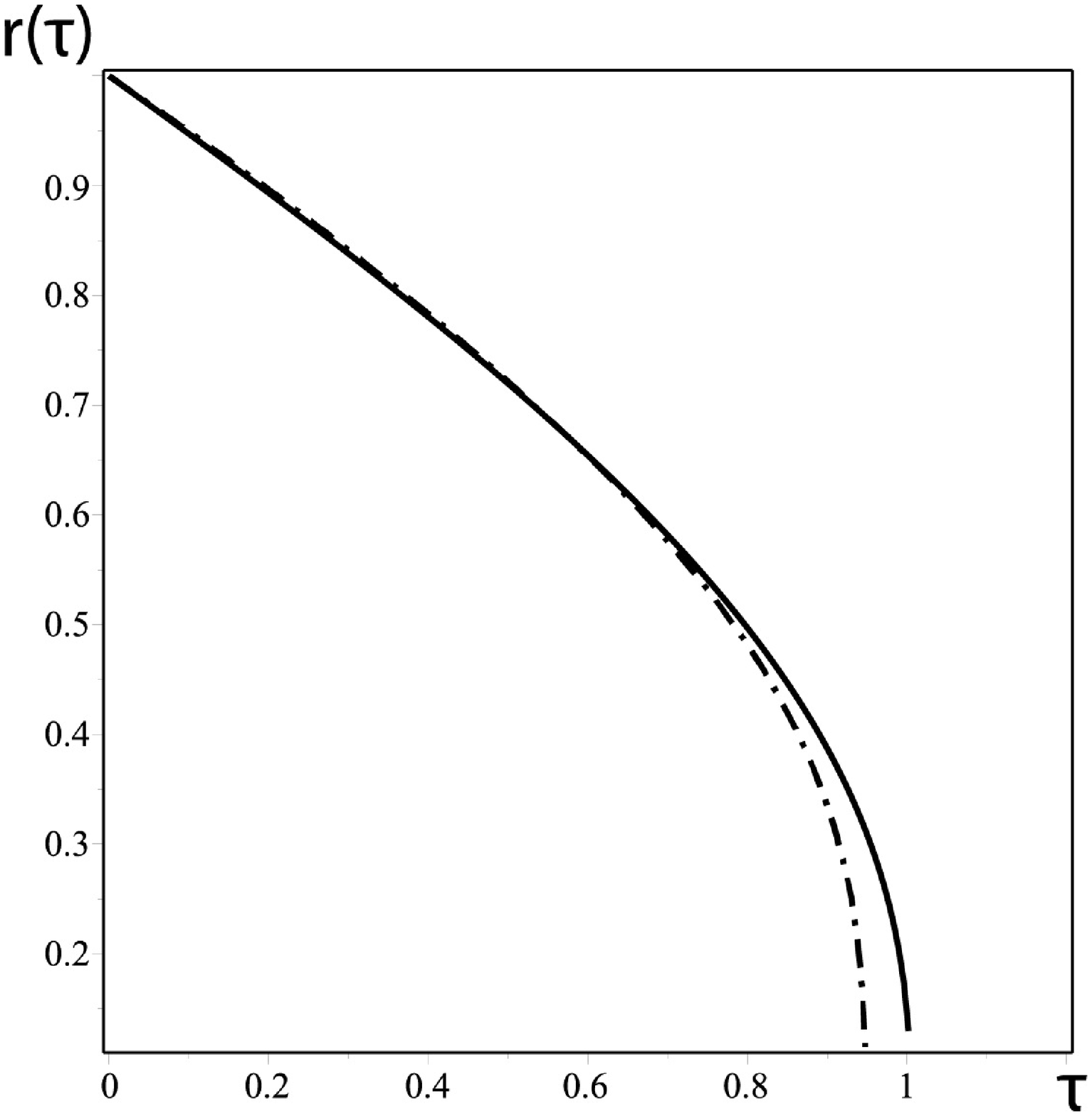}
	\includegraphics[height=7 cm]{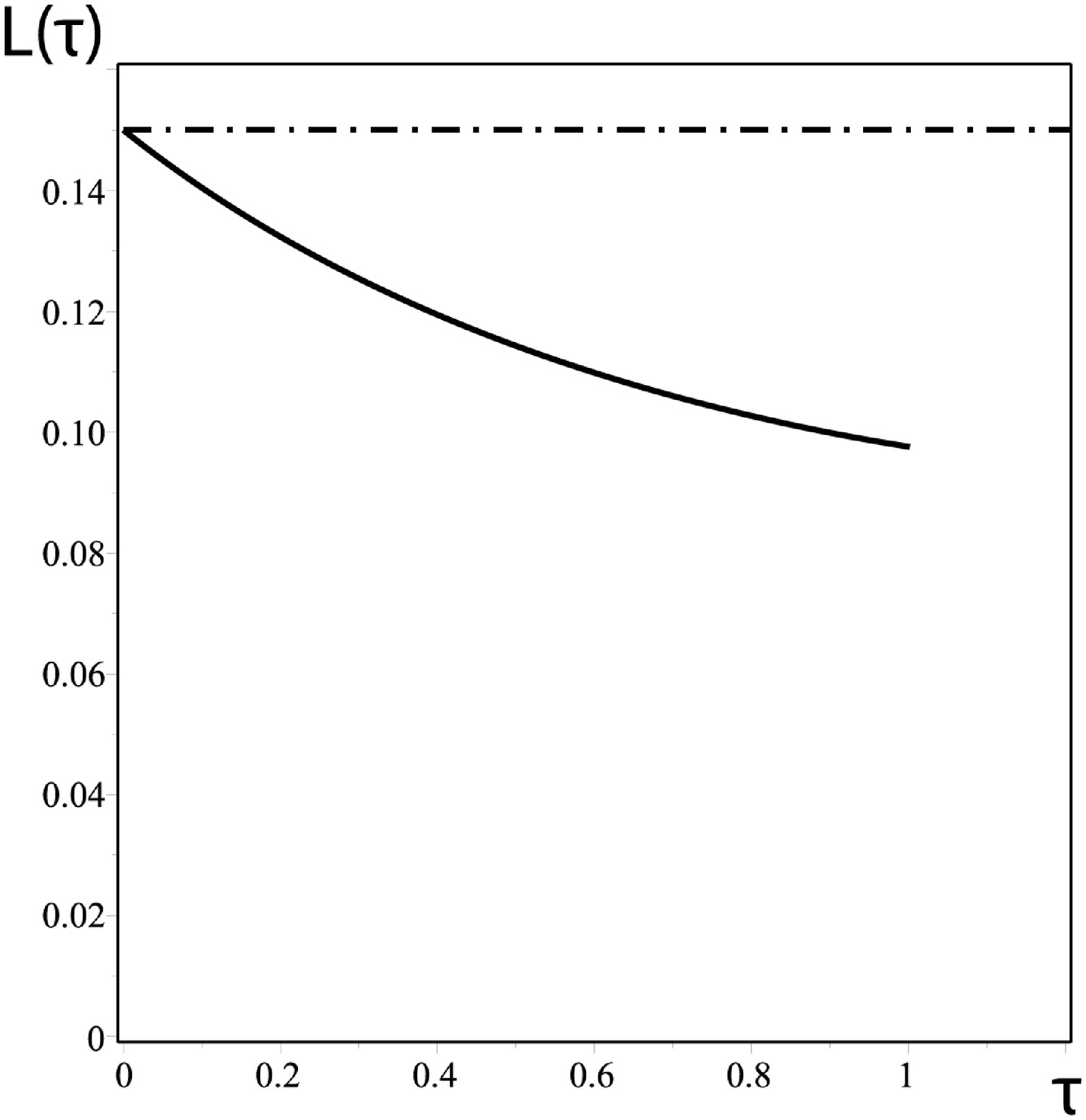}\\
	\caption{ On the left, the plots are shown of the pore radius time dependence: solid line designates the numerical solution of equation set (\ref{eq49}), dash-and-dot line corresponds to the solution of Eqs. (\ref{eq56})-(\ref{eq58}); on the right, time dependence for distance $L(\tau )$ is presented: solid line designates the numerical solution of Eqs. (\ref{eq49}), dash-and-dot line relates to the numerical solution of Eqs. (\ref{eq56})-(\ref{eq58}). All solutions are obtained for initial conditions $r|_{\tau =0} =1$, $r_{s} |_{\tau =0} =100$, $L|_{\tau =0} =90$ and $A=10^{-1} $.}
\label{fg10}
\end{figure}

Let us consider the valid regime of large pore evolution when relations between values $R$, $R_s $, $l$ are the following:
\begin{equation}\label{eq47}
 4) \quad R/R_s \cong 1,\quad l/R_s \ll 1,\quad  R \gg l. \end{equation}
 Let us write down the first relation as $R/R_{s} =1-\epsilon $, where $\epsilon $ is a small parameter of asymptotic expansion. With an account of the validity of conservation low for the volume of granule material, we can find, from Eq. (\ref{eq19}), granule radius change
 \[R_s(t)=\left(R_s(0)^3-R(0)^3+R(t)^3 \right)^{1/3} \]
or, in dimensionless units,
\begin{equation}\label{eq48} r_s(t)=\left(r_s(0)^3-r(0)^3+r(t)^3 \right)^{1/3} \end{equation}
Using this relation, we can describe the large pore evolution by the following dimensionless equations:
\begin{equation}\label{eq49}
\begin{cases}
  \frac{d L}{d \tau}=\frac{3\exp\left(\frac{A}{r}\right)}{r} \cdot \left[\frac{\alpha^2}{r^2} \cdot (\widetilde \Phi _1  + \widetilde \Phi _2 ) - \frac{\alpha}{r} \cdot \sqrt {1 + \frac{\alpha^2}{r^2}}  \cdot (\Phi _1  + \Phi _2 )\right]-\\
-\frac{6\exp\left(-\frac{A}{r_s(t)}\right)}{r}\cdot \left[\frac{\alpha^2}{r^2} \cdot \widetilde \Phi _2  - \frac{\alpha}{r} \cdot \sqrt {1 + \frac{\alpha^2}{r^2}}  \cdot \Phi _2 \right],\\
\frac{d r}{d \tau}=-\frac{\exp\left(\frac{A}{r}\right)}{r} \cdot \left[\frac{1}{2} + \frac{\alpha}{r} \cdot (\Phi _1  + \Phi _2 )\right]+\frac{2\exp\left(-\frac{A}{r_s(t)}\right)}{r}\cdot \frac{\alpha}{r} \cdot \Phi _2,\\
r_s(t)=\left(r_s(0)^3-r(0)^3+r(t)^3 \right)^{1/3}, \\
  r|_{\tau =0}=1,\\
  L|_{\tau =0}=\frac{l(0)}{R(0)}.
\end{cases}
\end{equation}
where
\[\alpha = \frac{r_{s}^2}{2L}\sqrt{1+\left(\frac{L^2}{r_{s}^2}-\frac{r^2}{r_{s}^2}\right)^2-2\left(\frac{L^2}{r_{s}^2}+\frac{r^2}{r_{s}^2}\right)}, \quad r_s=r_s(t).\]
The numerical solution of equation set (\ref{eq49}) is shown in Fig. \ref{fg10} by a solid line for initial data $r|_{\tau =0} =1$, $r_{s} |_{\tau =0} =1.5$, $L|_{\tau =0} =0.15$ and $A=10^{-1} $.

Let us consider asymptotic mode (\ref{eq47}) for a large pore, confining ourselves, in connection with Eq. (\ref{eq48}), to second-order terms on $\varepsilon $, that is
\begin{equation}\label{eq50} \epsilon(1-\epsilon)=\frac{V}{3r_s^3}, \end{equation}
 where $V=r_s(0)^3 -r(0)^3 $ is initial volume of the material. Substituting the value of $\epsilon =1-r/r_s $ into Eq. (\ref{eq50}),  we obtain quadratic equation for granule radius $r_s $, with the solution in the following form:
\begin{equation}\label{eq51} r_s=r\left(1+\frac{V}{3r^3}\right)  \end{equation}

Thus, within asymptotic approximation (\ref{eq47}), the connection is obtained between the pore and granule radii  (\ref{eq47}). Let us now proceed to the calculation of parameter $\alpha $, taking into account the condition $r \gg L$:
\begin{equation}\label{eq52} \alpha\approx\frac{r_s^2}{2L}\left(1+(1-\epsilon)^4-2(1-\epsilon)^2\right)^{1/2}=\frac{r_s^2\epsilon}{L}  \end{equation}
Hence, according to the definition (\ref{eq5}), one finds bispherical coordinates $\eta_{1,2} $:
\begin{equation}\label{eq53} \quad \eta _1  = \textrm{arsinh}\left(\frac{r_s^2\epsilon}{rL}\right),\quad
\eta _2 = \textrm{arsinh}\left(\frac{r_s}{L}\epsilon\right) \end{equation}
Because of the geometrical conditions, the inequality $\varepsilon r_s /L \geq 1$ is valid. Thus, bispherical coordinates $\eta_{1,2} $ can be approximated for the case $\varepsilon r_s /L  \gg 1$ in the following form:
\begin{equation}\label{eq54}
\eta _1  \approx \ln \left(\frac{2r_s^2\epsilon}{rL}\right),\quad \eta _2  \approx \ln \left(\frac{2r_s\epsilon}{L}\right) \end{equation}
Then, let us find the difference $\eta_1 -\eta_2 =\ln \left(\frac{r_s}{r} \right)\approx \ln (1+\epsilon )\approx \epsilon $ and, correspondingly, make an estimate of series sums:
\[ \Phi_1 =\frac{1}{2\sinh(\eta_1+\epsilon)} \approx \frac{1}{2(\sinh\eta_1+\epsilon\cosh\eta_1) }\approx\frac{r}{2\alpha(1+\epsilon)}=\frac{r}{2\alpha}\left(1-\epsilon+\epsilon^2+\cdots\right),\]
\begin{equation}\label{eq55}
 \Phi_2 =\frac{1}{2\sinh(\eta_2+\epsilon)} \approx  \frac{1}{2(\sinh\eta_2+\epsilon\cosh\eta_2) }\approx \frac{r_s}{2\alpha(1+\epsilon)}=\frac{r_s}{2\alpha}\left(1-\epsilon+\epsilon^2+\cdots\right),
\end{equation}
\[ \widetilde{\Phi}_1 = \frac{\cosh(\eta_1+\epsilon)}{2\sinh^2(\eta_1+\epsilon)} \approx \frac{\cosh\eta_1+\epsilon\sinh\eta_1 }{2(\sinh\eta_1+\epsilon\cosh\eta_1)^2 } \approx \frac{r}{2\alpha(1+\epsilon)}= \frac{r}{2\alpha}\left(1-\epsilon+\epsilon^2+\cdots\right),\]
\[ \widetilde{\Phi}_2 =\frac{\cosh(\eta_2+\epsilon)}{2\sinh^2(\eta_2+\epsilon)} \approx \frac{\cosh\eta_2+\epsilon\sinh\eta_2 }{2(\sinh\eta_2+\epsilon\cosh\eta_2)^2 } \approx \frac{r_s}{2\alpha(1+\epsilon)}=\frac{r_s}{2\alpha}\left(1-\epsilon+\epsilon^2+\cdots\right). \]
Let us substitute relations (\ref{eq51}), (\ref{eq52}) and (\ref{eq55}) into equation set (\ref{eq49}). Now we can obtain evolution equations for  the large pore with the accuracy up to second order term $\epsilon^2 $:
\begin{equation}\label{eq56} \frac{dL}{d\tau}=O(\epsilon^3) \end{equation}
\begin{equation}\label{eq57} \frac{dr}{d\tau}=-\frac{\exp\left(\frac{A}{r}\right)}{r}\cdot\left[\frac{1}{2}+\frac{1}{2}\cdot\left(\frac{2-\epsilon}{1-\epsilon}\right)(1-\epsilon+\epsilon^2)\right]+\frac{\exp\left(-\frac{A}{r\left(1+\frac{V}{3r^3}\right)}\right)}{r}\cdot\frac{1-\epsilon+\epsilon^2}{1-\epsilon} \end{equation}
It follows from Eq. (\ref{eq56}), that, within the considered asymptotic approximation, the change of distance $L(\tau )$ is quite small: $L(\tau )\approx L(0)$. Eq.(\ref{eq57}) does not depend on $L(\tau )$, thus, substituting into it the value $\epsilon =1-r/r_{s} $, we find pore radius evolution equation:
\begin{equation}\label{eq58}	 \frac{dr}{d\tau}=-\frac{\exp\left(\frac{A}{r}\right)}{r}\cdot\left[1+\frac{1}{2}\cdot\frac{1}{1+\frac{V}{3r^3}} +\left(\frac{\frac{V}{3r^3}}{1+\frac{V}{3r^3}}\right)^2 \right]+\frac{\exp\left(-\frac{A}{r\left(1+\frac{V}{3r^3}\right)}\right)}{r}\cdot\left(1+\left(\frac{\frac{V}{3r^3}}{1+\frac{V}{3r^3}}\right)^2\right) \end{equation}
 In Fig. \ref{fg10}, dash-and-dot line represents numerical solution of approximate equations (\ref{eq56})-(\ref{eq58}) with initial conditions $r|_{\tau =0} =1$, $r_{s} |_{\tau =0} =1.5$, $L|_{\tau =0} =0.15$ and $A=10^{-1} $. This figure demonstrates, that numerical solutions of exact (\ref{eq49}) and approximate  equations (\ref{eq56})-(\ref{eq58})  well agree. Even the discrepancies in the evolution of $L(\tau )$, observed in Fig. \ref{fg10}, are small (of the order of $\epsilon^3 $).  This corresponds to the next order of smallness, that was neglected when obtaining approximate Eqs. (\ref{eq56})-(\ref{eq58}).

\section{Conclusion}

Thus, the principal distinction of pore evolution in spherical granules consists in the absence of critical size that separates pore evolution modes. In an unbounded matrix, pores of the size larger than the critical one grow, while those of the smaller size dissolve. Pores inside spherical granules always diminish with time and move towards the granule center. The above said follows from the analysis of the obtained equation set. In general case, the pore is dissolving via vacancy mechanism before reaching granule center. Simple pore behavior is observed in the limiting cases of small and large pores. The character of pore volume decrease in the case of small pores, positioned at a short (of the order of pore size) distance from granule center, corresponds to linear time dependence. In the case of small pores, situated close to the granule boundary, the rate of pore diminishing turns to be proportional to square root of time. The analysis of limiting cases of small pores revealed explosive mode of pore size diminishing when reaching some small size. In all limiting cases, simple relations determining healing time of a pore inside the granule have been obtained.  Of course, qualitatively, the obtained results will hold for deviations of the pore shape from spherical one. Significant difference can occur for strongly anisotropic granules.

Of course, we did not take into account some additional complicating factors, that can reveal themselves in various cases. In principle, they can be taken into account at the cost of making the derivation of the equation system even more cumbersome. In particular, the majority of the effects produced by the various mechanisms of vacancy transport can be reduced to renormalizing of numerical coefficients of the obtained equation system.  A good example can be given by heterodiffusion in crystals of $Na Cl$ type, where vacancy transition between sublattices is practically impossible. This is caused by a great value of displacement energy. Then, vacancy flux is added up by concordant fluxes in each sublattice. Account of this process simply leads to renormalizing of the diffusion coefficient. Second factor, that is not considered within hydrodynamic approximation, is possible appearance of faceting for very small pores. However, one can expect, that the contribution due to influence of the shape of small  pore takes up a small part of the pore evolution type and, in the absence of strong anisotropy, is insignificant.

The considered simple case is important for the comparison of pore behavior in nanogranulas with numerical modeling results and for revealing general regularities of pore behavior.  Besides, it is useful for establishing coincidences and differences between the hydrodynamic approximation of pore behavior and the numerical modeling results. Such comparison is especially useful for establishing the applicability range of hydrodynamic approximation.

\section{Appendix}

\subsection{Auxiliary relations.}

\begin{equation}\label{EQ1}
\int_{-1}^1\frac{P_k(t)dt}{\sqrt{\cosh\eta-t}} = \frac{\sqrt{2}\cdot
e^{-(k+1/2)\eta}}{k+1/2}\,.
\end{equation}
Differentiating relation (\ref{EQ1}) with respect to parameter $\eta $,  one sequentially finds
\begin{equation}\label{EQ2}
\int_{-1}^1\frac{P_k(t)dt}{(\cosh\eta-t)^{3/2}} =
\frac{2\sqrt{2}\cdot e^{-(k+1/2)\eta}}{\sinh\eta}\,,
\end{equation}
\begin{equation}\label{EQ3}
\int_{-1}^1\frac{P_k(t)dt}{(\cosh\eta-t)^{5/2}} =
\frac{4\sqrt{2}\cdot
e^{-(k+1/2)\eta}(\cosh\eta+(k+1/2)\sinh\eta)}{3\cdot\sinh^3\eta}\,.
\end{equation}

\subsection{ The calculation of pore radius.}

\begin{equation}\label{EQ4} \dot{R}=-\frac{\omega}{4\pi
R^2}\oint\vec{n}\vec{j}dS\,,
\end{equation}
 where
\begin{equation}\label{EQ5}
\vec{n}\vec{j}|_{\eta=\eta_1}=\frac {D}{\omega} \cdot
\frac{\cosh\eta_1-\cos\xi}{a}\frac{\partial
c}{\partial\eta}|_{\eta=\eta_1}\,,
\end{equation}
 \begin{equation}\label{EQ6}
 dS = \frac{a^2\cdot \sin\xi
d\xi d\varphi}{(\cosh\eta_1-\cos\xi)^2}\,,
\end{equation}
According to the Fubini's theorem, due to independence of variables $\xi $ and $\varphi $, the expression for $\dot{R}$ takes on the form
 \begin{equation}\label{EQ7}
  \dot{R}=-\frac{a\cdot D}{2\cdot
R^2}\int_0^{\pi}\frac{\partial
c}{\partial\eta}|_{\eta=\eta_1}\frac{\sin\xi d\xi
}{\cosh\eta_1-\cos\xi}\,,
\end{equation}
Substituting
$$\frac{\partial
c}{\partial\eta}|_{\eta=\eta_1} = \sqrt{2}\left(
\frac{c_{R}\cdot\sinh\eta_1}{\sqrt{{\cosh\eta_1-\cos\xi}}}\cdot\sum_{k=0}^{\infty}
P_k(\cos\xi)\exp(-\eta_1(k+1/2))+
\sqrt{{\cosh\eta_1-\cos\xi}}\times\right.$$
$$\left.\times\sum_{k=0}^\infty\frac{(k+1/2)\cdot
P_k(\cos\xi)}{\sinh(k+1/2)(\eta_1-\eta_2)}\left[c_{R}\cdot\cosh(k+1/2)(\eta_1-\eta_2)e^{-\eta_1(k+1/2)}
-c_{R_s}\cdot e^{-\eta_2(k+1/2)}\right] \right)$$
into the expression for the rate  of pore radius change and transposing signs of summation and integration on the strength of the convergence of corresponding sums and integrals, after change of variable $\cos \xi =t$, one obtains
$$ \dot{R}=-\frac{a\cdot D\sqrt{2}}{2\cdot
R^2}\left[\frac{c_{R}\cdot\sinh\eta_1}{2}\sum_{k=0}^\infty
e^{-\eta_1(k+1)}\int_{-1}^1\frac{P_k(t)dt}{(\cosh\eta_1-t)^{3/2}}+\right.$$
$$\left. +\sum_{k=0}^\infty\frac{(k+1/2)}{\sinh(k+1/2)(\eta_1-\eta_2)}\left[c_{R}\cdot\cosh(k+1/2)
(\eta_1-\eta_2)e^{-\eta_1(k+1/2)} -\right.\right. $$
$$\left.\left.- c_{R_s}\cdot
e^{-\eta_2(k+1/2)}\right]\cdot
\int_{-1}^1\frac{P_k(t)dt}{\sqrt{\cosh\eta_1-t}}\right] \,.$$
Let us transform this expression with the help of integrals (\ref{EQ1}) and (\ref{EQ2}):
$$ \dot{R}=-\frac{a\cdot D\sqrt{2}}{2\cdot
R^2}\left[\frac{c_{R}\cdot\sinh\eta_1}{2}\sum_{k=0}^\infty
e^{-\eta_1(k+1/2)}\cdot \frac{2\sqrt{2}\cdot
e^{-\eta_1(k+1/2)}}{\sinh\eta_1} +\right.$$
$$\left. +\sum_{k=0}^\infty\frac{(k+1/2)}{\sinh(k+1/2)(\eta_1-\eta_2)}\left[c_{R}\cdot\cosh(k+1/2)
(\eta_1-\eta_2)e^{-\eta_1(k+1/2)} -\right.\right. $$
$$\left.\left.- c_{R_s}\cdot
e^{-\eta_2(k+1/2)}\right]\cdot \frac{\sqrt{2}\cdot
e^{-(k+1/2)\eta_1}}{k+1/2} \right]= -\frac{a\cdot D}{R^2}\left[
c_{R}\cdot\sum_{k=0}^\infty e^{-\eta_1(2k+1)} +\right.$$
$$\left. +\sum_{k=0}^\infty \frac{c_{R}\cdot\cosh(k+1/2)
(\eta_1-\eta_2)e^{-\eta_1(k+1/2)} + c_{R_s}\cdot e^{-\eta_2(k+1/2)}}
{\sinh(k+1/2)(\eta_1-\eta_2)}\cdot e^{-\eta_1(k+1/2)}\right] =
-\frac{a\cdot D}{R^2}\times$$
$$\times\left[\frac{c_{1}}{2\cdot\sinh\eta_1} +\sum_{k=0}^\infty
\frac{c_{R}\cdot\cosh(k+1/2) (\eta_1-\eta_2)e^{-\eta_1(k+1/2)} -
c_{R_s}\cdot e^{-\eta_2(k+1/2)}} {\sinh(k+1/2)(\eta_1-\eta_2)}\cdot
e^{-\eta_1(k+1/2)}\right]\,.$$
Substituting $a=R\cdot \sinh \eta _{1} $ and, reorganizing the expressions following summation symbols, one ultimately obtains:
\begin{equation}\label{EQ8} \dot{R}=-\frac{D}{R}\left[\frac{c_{R}}{2}
+\sinh\eta_1\cdot\sum_{k=0}^\infty
\frac{c_{R}\cdot(e^{-(2k+1)\eta_1}+e^{-(2k+1)\eta_2}) -2\cdot
c_{R_s}\cdot e^{-(2k+1)\eta_2}}
{e^{(2k+1)(\eta_1-\eta_2)}-1}\right]\,.
\end{equation}

\subsection{ The calculation of pore velocity.}

$$ \vec{v}=\vec{e_z}\cdot\frac{3\cdot D\cdot
a}{2\cdot R^2}\int_0^\pi \frac{\partial
c}{\partial\eta}|_{\eta=\eta_1}\frac{\cosh\eta_1\cdot
\cos\xi-1}{(\cosh\eta_1-\cos\xi)^2}\cdot\sin\xi d\xi = $$
$$=\vec{e_z}\cdot\frac{3\cdot D\cdot
a}{2\cdot R^2}\int_0^\pi \frac{\partial
c}{\partial\eta}|_{\eta=\eta_1}\left(-\frac{\cosh\eta_1}{\cosh\eta_1-\cos\xi}+
\frac{\sinh^2\eta_1}{(\cosh\eta_1-\cos\xi)^2}\right) \cdot\sin\xi
d\xi =$$
$$=\vec{e_z}\cdot\frac{3\sqrt{2}\cdot D\cdot
a}{2\cdot R^2}
\int_0^\pi\left(-\frac{\cosh\eta_1}{\cosh\eta_1-\cos\xi}+
\frac{\sinh^2\eta_1}{(\cosh\eta_1-\cos\xi)^2}\right) \cdot\sin\xi
d\xi\times$$
$$\times
\left[ \frac{c_R\cdot\sinh\eta_1}{2\cdot\sqrt{\cosh\eta_1-\cos\xi}}\cdot\sum_{k=0}^\infty P_k(\cos\xi)
e^{-\eta_1(k+1/2)} + \sqrt{\cosh\eta_1-\cos\xi}\times\right.$$

$$\left.\times\left( \sum_{k=0}^\infty\frac{(k+1/2)
P_k(\cos\xi)(c_R\cdot e^{-\eta_1(k+1/2)}\cosh(k+1/2)(\eta_1-\eta_2) -c_{R_s}\cdot
e^{-\eta_2(k+1/2)})}{\sinh(k+1/2)(\eta_1-\eta_2)}
 \right)\right]\,.$$
Substitution $t=\cos \xi $ and change of summation and integration order lead us to the expression
$$\vec{v}=\vec{e_z}\cdot\frac{3\sqrt{2}\cdot D\cdot
a}{2\cdot R^2}
\sum_{k=0}^\infty\left[\int_{-1}^1\frac{P_k(t)dt}{(\cosh\eta_1-t)^{5/2}}\cdot
\frac{c_R\cdot e^{-\eta_1(k+1/2)}\sinh^3\eta_1}{2}
\right.+$$
$$+\int_{-1}^1\frac{P_k(t)dt}{(\cosh\eta_1-t)^{3/2}}\times\left(\frac{-c_R\cdot e^{-\eta_1(k+1/2)}
\cosh\eta_1\sinh\eta_1}{2}\right.+\sinh^2\eta_1\times$$
$$\left.
\times\frac{(k+1/2)
(c_R\cdot e^{-\eta_1(k+1/2)}\cosh(k+1/2)(\eta_1-\eta_2) -c_{R_s}\cdot
e^{-\eta_2(k+1/2)})}{\sinh(k+1/2)(\eta_1-\eta_2)}
 \right)+\int_{-1}^1\frac{P_k(t)dt}{\sqrt{\cosh\eta_1-t}}\times$$
$$ \left.\times\left(
-\cosh\eta_1\cdot\frac{(k+1/2)
(c_R\cdot e^{-\eta_1(k+1/2)}\cosh(k+1/2)(\eta_1-\eta_2) -c_{R_s}\cdot
e^{-\eta_2(k+1/2)})}{\sinh(k+1/2)(\eta_1-\eta_2)}
 \right)\right]\,.$$
Substituting into the obtained expression the values of corresponding integrals, let us transform the result
$$\vec{v}=\vec{e_z}\cdot\frac{3\sqrt{2}\cdot D\cdot
a}{2\cdot R^2} \sum_{k=0}^\infty\left[\frac{4\sqrt{2}\cdot
e^{-(k+1/2)\eta_1}(\cosh\eta_1+(k+1/2)\sinh\eta_1)}{3\cdot\sinh^3\eta_1}\times
\right. $$
 $$\times\frac{c_R\cdot e^{-\eta_1(k+1/2)}\sinh^3\eta_1}{2}+\frac{2\sqrt{2}\cdot
e^{-(k+1/2)\eta_1}}{\sinh\eta_1}\times
\left(\frac{-c_R\cdot e^{-\eta_1(k+1/2)}\cosh\eta_1}{2}
+\sinh^2\eta_1\times\right.$$
$$\left. \times\frac{(k+1/2)
(c_R\cdot e^{-\eta_1(k+1/2)}\cosh(k+1/2)(\eta_1-\eta_2) -c_{R_s}\cdot
e^{-\eta_2(k+1/2)})}{\sinh(k+1/2)(\eta_1-\eta_2)}
 \right)+\frac{\sqrt{2}\cdot e^{-(k+1/2)\eta_1}}{k+1/2}\times$$
$$\left. \times\left(
-\cosh\eta_1\cdot\frac{(k+1/2)
(c_R \cdot e^{-\eta_1(k+1/2)}\cosh(k+1/2)(\eta_1-\eta_2) -c_{R_s}\cdot
e^{-\eta_2(k+1/2)})}{\sinh(k+1/2)(\eta_1-\eta_2)}
 \right)\right]=$$

\begin{equation}\label{EQ9}=\vec{e_z}\cdot\frac{3 D
a}{ R^2}
\sum_{k=0}^\infty\frac{((2k+1)\sinh\eta_1-\cosh\eta_1)\left(c_R\cdot(e^{-(2k+1)\eta_1}+e^{-(2k+1)\eta_2})-2c_{R_s}\cdot
e^{-(2k+1)\eta_2}\right)}{e^{(2k+1)(\eta_1-\eta_2)}-1}\,.
\end{equation}

\end{document}